\documentclass[twocolumn]{aastex63}

\usepackage[utf8]{inputenc}
\usepackage{enumitem}
\usepackage{amsmath}
\usepackage{graphicx}
\usepackage{hyperref}
\usepackage{color}
\hypersetup{colorlinks=true, citecolor=cyan, urlcolor = blue, linkcolor = blue}
\usepackage{multirow}

\newcommand{\cc}{$\checkmark$}

\shorttitle{NANOGrav 15-year Individual Binary Limits}
\shortauthors{The NANOGrav Collaboration}

\begin{document}
\title{The NANOGrav 15-year Data Set: Bayesian Limits on Gravitational Waves from Individual Supermassive Black Hole Binaries}

\author[0000-0001-5134-3925]{Gabriella Agazie}
\affiliation{Center for Gravitation, Cosmology and Astrophysics, Department of Physics, University of Wisconsin-Milwaukee,\\ P.O. Box 413, Milwaukee, WI 53201, USA}
\author[0000-0002-8935-9882]{Akash Anumarlapudi}
\affiliation{Center for Gravitation, Cosmology and Astrophysics, Department of Physics, University of Wisconsin-Milwaukee,\\ P.O. Box 413, Milwaukee, WI 53201, USA}
\author[0000-0003-0638-3340]{Anne M. Archibald}
\affiliation{Newcastle University, NE1 7RU, UK}
\author{Zaven Arzoumanian}
\affiliation{X-Ray Astrophysics Laboratory, NASA Goddard Space Flight Center, Code 662, Greenbelt, MD 20771, USA}
\author[0000-0003-2745-753X]{Paul T. Baker}
\affiliation{Department of Physics and Astronomy, Widener University, One University Place, Chester, PA 19013, USA}
\author[0000-0003-0909-5563]{Bence B\'{e}csy}
\affiliation{Department of Physics, Oregon State University, Corvallis, OR 97331, USA}
\author[0000-0002-2183-1087]{Laura Blecha}
\affiliation{Physics Department, University of Florida, Gainesville, FL 32611, USA}
\author[0000-0001-6341-7178]{Adam Brazier}
\affiliation{Cornell Center for Astrophysics and Planetary Science and Department of Astronomy, Cornell University, Ithaca, NY 14853, USA}
\affiliation{Cornell Center for Advanced Computing, Cornell University, Ithaca, NY 14853, USA}
\author[0000-0003-3053-6538]{Paul R. Brook}
\affiliation{Institute for Gravitational Wave Astronomy and School of Physics and Astronomy, University of Birmingham, Edgbaston, Birmingham B15 2TT, UK}
\author[0000-0003-4052-7838]{Sarah Burke-Spolaor}
\affiliation{Department of Physics and Astronomy, West Virginia University, P.O. Box 6315, Morgantown, WV 26506, USA}
\affiliation{Center for Gravitational Waves and Cosmology, West Virginia University, Chestnut Ridge Research Building, Morgantown, WV 26505, USA}
\author{Robin Case}
\affiliation{Department of Physics, Oregon State University, Corvallis, OR 97331, USA}
\author[0000-0002-5557-4007]{J. Andrew Casey-Clyde}
\affiliation{Department of Physics, University of Connecticut, 196 Auditorium Road, U-3046, Storrs, CT 06269-3046, USA}
\author[0000-0003-3579-2522]{Maria Charisi}
\affiliation{Department of Physics and Astronomy, Vanderbilt University, 2301 Vanderbilt Place, Nashville, TN 37235, USA}
\author[0000-0002-2878-1502]{Shami Chatterjee}
\affiliation{Cornell Center for Astrophysics and Planetary Science and Department of Astronomy, Cornell University, Ithaca, NY 14853, USA}
\author[0000-0001-7587-5483]{Tyler Cohen}
\affiliation{Department of Physics, New Mexico Institute of Mining and Technology, 801 Leroy Place, Socorro, NM 87801, USA}
\author[0000-0002-4049-1882]{James M. Cordes}
\affiliation{Cornell Center for Astrophysics and Planetary Science and Department of Astronomy, Cornell University, Ithaca, NY 14853, USA}
\author[0000-0002-7435-0869]{Neil J. Cornish}
\affiliation{Department of Physics, Montana State University, Bozeman, MT 59717, USA}
\author[0000-0002-2578-0360]{Fronefield Crawford}
\affiliation{Department of Physics and Astronomy, Franklin \& Marshall College, P.O. Box 3003, Lancaster, PA 17604, USA}
\author[0000-0002-6039-692X]{H. Thankful Cromartie}
\altaffiliation{NASA Hubble Fellowship: Einstein Postdoctoral Fellow}
\affiliation{Cornell Center for Astrophysics and Planetary Science and Department of Astronomy, Cornell University, Ithaca, NY 14853, USA}
\author[0000-0002-1529-5169]{Kathryn Crowter}
\affiliation{Department of Physics and Astronomy, University of British Columbia, 6224 Agricultural Road, Vancouver, BC V6T 1Z1, Canada}
\author[0000-0002-2185-1790]{Megan E. DeCesar}
\affiliation{George Mason University, resident at the Naval Research Laboratory, Washington, DC 20375, USA}
\author[0000-0002-6664-965X]{Paul B. Demorest}
\affiliation{National Radio Astronomy Observatory, 1003 Lopezville Rd., Socorro, NM 87801, USA}
\author{Matthew C. Digman}
\affiliation{Department of Physics, Montana State University, Bozeman, MT 59717, USA}
\author[0000-0001-8885-6388]{Timothy Dolch}
\affiliation{Department of Physics, Hillsdale College, 33 E. College Street, Hillsdale, MI 49242, USA}
\affiliation{Eureka Scientific, 2452 Delmer Street, Suite 100, Oakland, CA 94602-3017, USA}
\author{Brendan Drachler}
\affiliation{School of Physics and Astronomy, Rochester Institute of Technology, Rochester, NY 14623, USA}
\affiliation{Laboratory for Multiwavelength Astrophysics, Rochester Institute of Technology, Rochester, NY 14623, USA}
\author[0000-0001-7828-7708]{Elizabeth C. Ferrara}
\affiliation{Department of Astronomy, University of Maryland, College Park, MD 20742}
\affiliation{Center for Research and Exploration in Space Science and Technology, NASA/GSFC, Greenbelt, MD 20771}
\affiliation{NASA Goddard Space Flight Center, Greenbelt, MD 20771, USA}
\author[0000-0001-5645-5336]{William Fiore}
\affiliation{Department of Physics and Astronomy, West Virginia University, P.O. Box 6315, Morgantown, WV 26506, USA}
\affiliation{Center for Gravitational Waves and Cosmology, West Virginia University, Chestnut Ridge Research Building, Morgantown, WV 26505, USA}
\author[0000-0001-8384-5049]{Emmanuel Fonseca}
\affiliation{Department of Physics and Astronomy, West Virginia University, P.O. Box 6315, Morgantown, WV 26506, USA}
\affiliation{Center for Gravitational Waves and Cosmology, West Virginia University, Chestnut Ridge Research Building, Morgantown, WV 26505, USA}
\author[0000-0001-7624-4616]{Gabriel E. Freedman}
\affiliation{Center for Gravitation, Cosmology and Astrophysics, Department of Physics, University of Wisconsin-Milwaukee,\\ P.O. Box 413, Milwaukee, WI 53201, USA}
\author[0000-0001-6166-9646]{Nate Garver-Daniels}
\affiliation{Department of Physics and Astronomy, West Virginia University, P.O. Box 6315, Morgantown, WV 26506, USA}
\affiliation{Center for Gravitational Waves and Cosmology, West Virginia University, Chestnut Ridge Research Building, Morgantown, WV 26505, USA}
\author[0000-0001-8158-683X]{Peter A. Gentile}
\affiliation{Department of Physics and Astronomy, West Virginia University, P.O. Box 6315, Morgantown, WV 26506, USA}
\affiliation{Center for Gravitational Waves and Cosmology, West Virginia University, Chestnut Ridge Research Building, Morgantown, WV 26505, USA}
\author[0000-0003-4090-9780]{Joseph Glaser}
\affiliation{Department of Physics and Astronomy, West Virginia University, P.O. Box 6315, Morgantown, WV 26506, USA}
\affiliation{Center for Gravitational Waves and Cosmology, West Virginia University, Chestnut Ridge Research Building, Morgantown, WV 26505, USA}
\author[0000-0003-1884-348X]{Deborah C. Good}
\affiliation{Department of Physics, University of Connecticut, 196 Auditorium Road, U-3046, Storrs, CT 06269-3046, USA}
\affiliation{Center for Computational Astrophysics, Flatiron Institute, 162 5th Avenue, New York, NY 10010, USA}
\author[0000-0002-1146-0198]{Kayhan G\"{u}ltekin}
\affiliation{Department of Astronomy and Astrophysics, University of Michigan, Ann Arbor, MI 48109, USA}
\author[0000-0003-2742-3321]{Jeffrey S. Hazboun}
\affiliation{Department of Physics, Oregon State University, Corvallis, OR 97331, USA}
\author{Sophie Hourihane}
\affiliation{Division of Physics, Mathematics, and Astronomy, California Institute of Technology, Pasadena, CA 91125, USA}
\author[0000-0003-1082-2342]{Ross J. Jennings}
\altaffiliation{NANOGrav Physics Frontiers Center Postdoctoral Fellow}
\affiliation{Department of Physics and Astronomy, West Virginia University, P.O. Box 6315, Morgantown, WV 26506, USA}
\affiliation{Center for Gravitational Waves and Cosmology, West Virginia University, Chestnut Ridge Research Building, Morgantown, WV 26505, USA}
\author[0000-0002-7445-8423]{Aaron D. Johnson}
\affiliation{Center for Gravitation, Cosmology and Astrophysics, Department of Physics, University of Wisconsin-Milwaukee,\\ P.O. Box 413, Milwaukee, WI 53201, USA}
\affiliation{Division of Physics, Mathematics, and Astronomy, California Institute of Technology, Pasadena, CA 91125, USA}
\author[0000-0001-6607-3710]{Megan L. Jones}
\affiliation{Center for Gravitation, Cosmology and Astrophysics, Department of Physics, University of Wisconsin-Milwaukee,\\ P.O. Box 413, Milwaukee, WI 53201, USA}
\author[0000-0002-3654-980X]{Andrew R. Kaiser}
\affiliation{Department of Physics and Astronomy, West Virginia University, P.O. Box 6315, Morgantown, WV 26506, USA}
\affiliation{Center for Gravitational Waves and Cosmology, West Virginia University, Chestnut Ridge Research Building, Morgantown, WV 26505, USA}
\author[0000-0001-6295-2881]{David L. Kaplan}
\affiliation{Center for Gravitation, Cosmology and Astrophysics, Department of Physics, University of Wisconsin-Milwaukee,\\ P.O. Box 413, Milwaukee, WI 53201, USA}
\author[0000-0002-6625-6450]{Luke Zoltan Kelley}
\affiliation{Department of Astronomy, University of California, Berkeley, 501 Campbell Hall \#3411, Berkeley, CA 94720, USA}
\author[0000-0002-0893-4073]{Matthew Kerr}
\affiliation{Space Science Division, Naval Research Laboratory, Washington, DC 20375-5352, USA}
\author[0000-0003-0123-7600]{Joey S. Key}
\affiliation{University of Washington Bothell, 18115 Campus Way NE, Bothell, WA 98011, USA}
\author[0000-0002-9197-7604]{Nima Laal}
\affiliation{Department of Physics, Oregon State University, Corvallis, OR 97331, USA}
\author[0000-0003-0721-651X]{Michael T. Lam}
\affiliation{School of Physics and Astronomy, Rochester Institute of Technology, Rochester, NY 14623, USA}
\affiliation{Laboratory for Multiwavelength Astrophysics, Rochester Institute of Technology, Rochester, NY 14623, USA}
\author[0000-0003-1096-4156]{William G. Lamb}
\affiliation{Department of Physics and Astronomy, Vanderbilt University, 2301 Vanderbilt Place, Nashville, TN 37235, USA}
\author{T. Joseph W. Lazio}
\affiliation{Jet Propulsion Laboratory, California Institute of Technology, 4800 Oak Grove Drive, Pasadena, CA 91109, USA}
\author[0000-0003-0771-6581]{Natalia Lewandowska}
\affiliation{Department of Physics, State University of New York at Oswego, Oswego, NY, 13126, USA}
\author[0000-0001-5766-4287]{Tingting Liu}
\affiliation{Department of Physics and Astronomy, West Virginia University, P.O. Box 6315, Morgantown, WV 26506, USA}
\affiliation{Center for Gravitational Waves and Cosmology, West Virginia University, Chestnut Ridge Research Building, Morgantown, WV 26505, USA}
\author[0000-0003-1301-966X]{Duncan R. Lorimer}
\affiliation{Department of Physics and Astronomy, West Virginia University, P.O. Box 6315, Morgantown, WV 26506, USA}
\affiliation{Center for Gravitational Waves and Cosmology, West Virginia University, Chestnut Ridge Research Building, Morgantown, WV 26505, USA}
\author[0000-0001-5373-5914]{Jing Luo}
\altaffiliation{Deceased}
\affiliation{Department of Astronomy \& Astrophysics, University of Toronto, 50 Saint George Street, Toronto, ON M5S 3H4, Canada}
\author[0000-0001-5229-7430]{Ryan S. Lynch}
\affiliation{Green Bank Observatory, P.O. Box 2, Green Bank, WV 24944, USA}
\author[0000-0002-4430-102X]{Chung-Pei Ma}
\affiliation{Department of Astronomy, University of California, Berkeley, 501 Campbell Hall \#3411, Berkeley, CA 94720, USA}
\affiliation{Department of Physics, University of California, Berkeley, CA 94720, USA}
\author[0000-0003-2285-0404]{Dustin R. Madison}
\affiliation{Department of Physics, University of the Pacific, 3601 Pacific Avenue, Stockton, CA 95211, USA}
\author[0000-0001-5481-7559]{Alexander McEwen}
\affiliation{Center for Gravitation, Cosmology and Astrophysics, Department of Physics, University of Wisconsin-Milwaukee,\\ P.O. Box 413, Milwaukee, WI 53201, USA}
\author[0000-0002-2885-8485]{James W. McKee}
\affiliation{E.A. Milne Centre for Astrophysics, University of Hull, Cottingham Road, Kingston-upon-Hull, HU6 7RX, UK}
\affiliation{Centre of Excellence for Data Science, Artificial Intelligence and Modelling (DAIM), University of Hull, Cottingham Road, Kingston-upon-Hull, HU6 7RX, UK}
\author[0000-0001-7697-7422]{Maura A. McLaughlin}
\affiliation{Department of Physics and Astronomy, West Virginia University, P.O. Box 6315, Morgantown, WV 26506, USA}
\affiliation{Center for Gravitational Waves and Cosmology, West Virginia University, Chestnut Ridge Research Building, Morgantown, WV 26505, USA}
\author[0000-0002-4642-1260]{Natasha McMann}
\affiliation{Department of Physics and Astronomy, Vanderbilt University, 2301 Vanderbilt Place, Nashville, TN 37235, USA}
\author[0000-0001-8845-1225]{Bradley W. Meyers}
\affiliation{Department of Physics and Astronomy, University of British Columbia, 6224 Agricultural Road, Vancouver, BC V6T 1Z1, Canada}
\affiliation{International Centre for Radio Astronomy Research, Curtin University, Bentley, WA 6102, Australia}
\author[0000-0002-2689-0190]{Patrick M. Meyers}
\affiliation{Division of Physics, Mathematics, and Astronomy, California Institute of Technology, Pasadena, CA 91125, USA}
\author[0000-0002-4307-1322]{Chiara M. F. Mingarelli}
\affiliation{Center for Computational Astrophysics, Flatiron Institute, 162 5th Avenue, New York, NY 10010, USA}
\affiliation{Department of Physics, University of Connecticut, 196 Auditorium Road, U-3046, Storrs, CT 06269-3046, USA}
\affiliation{Department of Physics, Yale University, New Haven, CT 06520, USA}
\author[0000-0003-2898-5844]{Andrea Mitridate}
\affiliation{Deutsches Elektronen-Synchrotron DESY, Notkestr. 85, 22607 Hamburg, Germany}
\author[0000-0002-3616-5160]{Cherry Ng}
\affiliation{Dunlap Institute for Astronomy and Astrophysics, University of Toronto, 50 St. George St., Toronto, ON M5S 3H4, Canada}
\author[0000-0002-6709-2566]{David J. Nice}
\affiliation{Department of Physics, Lafayette College, Easton, PA 18042, USA}
\author[0000-0002-4941-5333]{Stella Koch Ocker}
\affiliation{Cornell Center for Astrophysics and Planetary Science and Department of Astronomy, Cornell University, Ithaca, NY 14853, USA}
\author[0000-0002-2027-3714]{Ken D. Olum}
\affiliation{Institute of Cosmology, Department of Physics and Astronomy, Tufts University, Medford, MA 02155, USA}
\author[0000-0001-5465-2889]{Timothy T. Pennucci}
\affiliation{Institute of Physics and Astronomy, E\"{o}tv\"{o}s Lor\'{a}nd University, P\'{a}zm\'{a}ny P. s. 1/A, 1117 Budapest, Hungary}
\author[0000-0002-8509-5947]{Benetge B. P. Perera}
\affiliation{Arecibo Observatory, HC3 Box 53995, Arecibo, PR 00612, USA}
\author[0000-0001-5681-4319]{Polina Petrov}
\affiliation{Department of Physics and Astronomy, Vanderbilt University, 2301 Vanderbilt Place, Nashville, TN 37235, USA}
\author[0000-0002-8826-1285]{Nihan S. Pol}
\affiliation{Department of Physics and Astronomy, Vanderbilt University, 2301 Vanderbilt Place, Nashville, TN 37235, USA}
\author[0000-0002-2074-4360]{Henri A. Radovan}
\affiliation{Department of Physics, University of Puerto Rico, Mayag\"{u}ez, PR 00681, USA}
\author[0000-0001-5799-9714]{Scott M. Ransom}
\affiliation{National Radio Astronomy Observatory, 520 Edgemont Road, Charlottesville, VA 22903, USA}
\author[0000-0002-5297-5278]{Paul S. Ray}
\affiliation{Space Science Division, Naval Research Laboratory, Washington, DC 20375-5352, USA}
\author[0000-0003-4915-3246]{Joseph D. Romano}
\affiliation{Department of Physics, Texas Tech University, Box 41051, Lubbock, TX 79409, USA}
\author[0009-0006-5476-3603]{Shashwat C. Sardesai}
\affiliation{Center for Gravitation, Cosmology and Astrophysics, Department of Physics, University of Wisconsin-Milwaukee,\\ P.O. Box 413, Milwaukee, WI 53201, USA}
\author[0000-0003-4391-936X]{Ann Schmiedekamp}
\affiliation{Department of Physics, Penn State Abington, Abington, PA 19001, USA}
\author[0000-0002-1283-2184]{Carl Schmiedekamp}
\affiliation{Department of Physics, Penn State Abington, Abington, PA 19001, USA}
\author[0000-0003-2807-6472]{Kai Schmitz}
\affiliation{Institute for Theoretical Physics, University of M\"{u}nster, 48149 M\"{u}nster, Germany}
\author[0000-0002-7283-1124]{Brent J. Shapiro-Albert}
\affiliation{Department of Physics and Astronomy, West Virginia University, P.O. Box 6315, Morgantown, WV 26506, USA}
\affiliation{Center for Gravitational Waves and Cosmology, West Virginia University, Chestnut Ridge Research Building, Morgantown, WV 26505, USA}
\affiliation{Giant Army, 915A 17th Ave, Seattle WA 98122}
\author[0000-0002-7778-2990]{Xavier Siemens}
\affiliation{Department of Physics, Oregon State University, Corvallis, OR 97331, USA}
\affiliation{Center for Gravitation, Cosmology and Astrophysics, Department of Physics, University of Wisconsin-Milwaukee,\\ P.O. Box 413, Milwaukee, WI 53201, USA}
\author[0000-0003-1407-6607]{Joseph Simon}
\altaffiliation{NSF Astronomy and Astrophysics Postdoctoral Fellow}
\affiliation{Department of Astrophysical and Planetary Sciences, University of Colorado, Boulder, CO 80309, USA}
\author[0000-0002-1530-9778]{Magdalena S. Siwek}
\affiliation{Center for Astrophysics, Harvard University, 60 Garden St, Cambridge, MA 02138}
\author[0000-0001-9784-8670]{Ingrid H. Stairs}
\affiliation{Department of Physics and Astronomy, University of British Columbia, 6224 Agricultural Road, Vancouver, BC V6T 1Z1, Canada}
\author[0000-0002-1797-3277]{Daniel R. Stinebring}
\affiliation{Department of Physics and Astronomy, Oberlin College, Oberlin, OH 44074, USA}
\author[0000-0002-7261-594X]{Kevin Stovall}
\affiliation{National Radio Astronomy Observatory, 1003 Lopezville Rd., Socorro, NM 87801, USA}
\author[0000-0002-2820-0931]{Abhimanyu Susobhanan}
\affiliation{Center for Gravitation, Cosmology and Astrophysics, Department of Physics, University of Wisconsin-Milwaukee,\\ P.O. Box 413, Milwaukee, WI 53201, USA}
\author[0000-0002-1075-3837]{Joseph K. Swiggum}
\altaffiliation{NANOGrav Physics Frontiers Center Postdoctoral Fellow}
\affiliation{Department of Physics, Lafayette College, Easton, PA 18042, USA}
\author{Jacob Taylor}
\affiliation{Department of Physics, Oregon State University, Corvallis, OR 97331, USA}
\author[0000-0003-0264-1453]{Stephen R. Taylor}
\affiliation{Department of Physics and Astronomy, Vanderbilt University, 2301 Vanderbilt Place, Nashville, TN 37235, USA}
\author[0000-0002-2451-7288]{Jacob E. Turner}
\affiliation{Department of Physics and Astronomy, West Virginia University, P.O. Box 6315, Morgantown, WV 26506, USA}
\affiliation{Center for Gravitational Waves and Cosmology, West Virginia University, Chestnut Ridge Research Building, Morgantown, WV 26505, USA}
\author[0000-0001-8800-0192]{Caner Unal}
\affiliation{Department of Physics, Ben-Gurion University of the Negev, Be'er Sheva 84105, Israel}
\affiliation{Feza Gursey Institute, Bogazici University, Kandilli, 34684, Istanbul, Turkey}
\author[0000-0002-4162-0033]{Michele Vallisneri}
\affiliation{Jet Propulsion Laboratory, California Institute of Technology, 4800 Oak Grove Drive, Pasadena, CA 91109, USA}
\affiliation{Division of Physics, Mathematics, and Astronomy, California Institute of Technology, Pasadena, CA 91125, USA}
\author[0000-0002-6428-2620]{Rutger van~Haasteren}
\affiliation{Max-Planck-Institut f\"{u}r Gravitationsphysik (Albert-Einstein-Institut), Callinstrasse 38, D-30167, Hannover, Germany}
\author[0000-0003-4700-9072]{Sarah J. Vigeland}
\affiliation{Center for Gravitation, Cosmology and Astrophysics, Department of Physics, University of Wisconsin-Milwaukee,\\ P.O. Box 413, Milwaukee, WI 53201, USA}
\author[0000-0001-9678-0299]{Haley M. Wahl}
\affiliation{Department of Physics and Astronomy, West Virginia University, P.O. Box 6315, Morgantown, WV 26506, USA}
\affiliation{Center for Gravitational Waves and Cosmology, West Virginia University, Chestnut Ridge Research Building, Morgantown, WV 26505, USA}
\author[0000-0002-6020-9274]{Caitlin A. Witt}
\affiliation{Center for Interdisciplinary Exploration and Research in Astrophysics (CIERA), Northwestern University, Evanston, IL 60208}
\affiliation{Adler Planetarium, 1300 S. DuSable Lake Shore Dr., Chicago, IL 60605, USA}
\author[0000-0002-0883-0688]{Olivia Young}
\affiliation{School of Physics and Astronomy, Rochester Institute of Technology, Rochester, NY 14623, USA}
\affiliation{Laboratory for Multiwavelength Astrophysics, Rochester Institute of Technology, Rochester, NY 14623, USA}

\collaboration{1000}{The NANOGrav Collaboration}
\noaffiliation

\correspondingauthor{The NANOGrav Collaboration}
\email{comments@nanograv.org}

\begin{abstract}
Evidence for a low-frequency stochastic gravitational wave background has recently been reported based on analyses of pulsar timing array data. The most likely source of such a background is a population of supermassive black hole binaries, the loudest of which may be individually detected in these datasets. Here we present the search for individual supermassive black hole binaries in the NANOGrav 15-year dataset. We introduce several new techniques, which enhance the efficiency and modeling accuracy of the analysis. The search uncovered weak evidence for two candidate signals, one with a gravitational-wave frequency of $\sim$4 nHz, and another at $\sim$170 nHz. The significance of the low-frequency candidate was greatly diminished when Hellings-Downs correlations were included in the background model. The high-frequency candidate was discounted due to the lack of a plausible host galaxy, the unlikely astrophysical prior odds of finding such a source, and since most of its support comes from a single pulsar with a commensurate binary period. Finding no compelling evidence for signals from individual binary systems, we place upper limits on the strain amplitude of gravitational waves emitted by such systems. 
At our most sensitive frequency of 6 nHz we place a sky-averaged 95\% upper limit of $8\times10^{-15}$ on the strain amplitude. We also calculate an exclusion volume and a corresponding effective radius, within which we can rule out the presence of black hole binaries emitting at a given frequency. 
\end{abstract}


\section{Introduction}
Nanohertz gravitational waves (GWs) can be probed by regularly monitoring millisecond pulsars and measuring the times-of-arrival of their radio pulses (for a review see e.g.,~\citealt{PTA_review, SteveBook}). Three such pulsar timing arrays (PTAs) have the required decade-long datasets to probe the nHz band: the North American Nanohertz Observatory for Gravitational Waves (NANOGrav - \citealt{nanograv_15yr_dataset}); the European Pulsar Timing Array (EPTA - \citealt{epta_dr2_dataset}); and the Parkes Pulsar Timing Array (PPTA - \citealt{ppta_dr3_dataset}).
All three of these collaborations found a low-frequency stochastic process in their data~\citep{nanograv_15yr_gwb, epta_dr2_gwb, ppta_dr3_gwb}. They also found various levels of evidence for Hellings-Downs (HD, \citealt{HD}) spatial correlations between pulsars, which points to the origin of this process being a stochastic gravitational-wave background (GWB).
The upcoming combination of these datasets, along with data from the Indian Pulsar Timing Array (InPTA \citealt{inpta_dr1_data}), will constitute the next dataset of the International PTA (IPTA), which is expected to be more conclusive than any individual PTA.

The canonical explanation of such a GWB is that it is built up from the combined signals from a collection of millions of supermassive black hole binaries (SMBHBs) throughout the observable Universe (see \citealt{nanograv_15yr_astro} and references therein). These systems naturally form in galaxy mergers, since every massive galaxy hosts a supermassive black hole \citep{Kormendy+2013}, with mass of $10^6-10^{10}M_{\odot}$. Their existence has been hypothesized for decades \citep{Begelman+1980}, but they have remained observationally elusive despite their expected abundance \citep{DeRosa+2019}. It is expected that some of the loudest binaries could also be individually detected \citep{SVV2009, rosado_expected_properties, Luke_single_source, Becsy:2022pnr}. Several previous searches have been carried out to look for these sources, which have set increasingly stringent upper limits over the years \citep{yardley+10, nanograv_5yr_cw, PPTA-cw-paper, EPTA-CW-paper, nanograv_11yr_cw, nanograv_12p5yr_cw, ipta_dr2_cw}, including stringent mass ratio upper limits on tentative binaries in several nearby galaxies \citep{Schutz+2016,Arzoumanian+2021}. Most recently, \citet{nanograv_12p5yr_cw} searched for individual binaries in the NANOGrav 12.5-year dataset \citep{nanograv_12p5yr_data}. 

Here we describe a Bayesian search for individual binaries in the NANOGrav 15-year dataset \citep{nanograv_15yr_dataset}. The plan of this paper is as follows. In Section \ref{sec:methods}, we describe the search methods, which are based upon those in \citet{nanograv_12p5yr_cw}, but to which we have introduced several improvements. We present our search results in Section \ref{sec:results} and upper limits in Section \ref{sec:limits}. We discuss conclusions and future work in Section \ref{sec:conclusion}.

\section{Methods}
\label{sec:methods}

\subsection{The NANOGrav 15-year Data Set}
We analyze the NANOGrav 15-year dataset\footnote{While the time between the first and last observations we analyze is 16.03 years, this data set is named “15-year data set” since no single pulsar exceeds 16 years of observation; we will use this nomenclature despite the discrepancy.}, which consists of times-of-arrival and timing models of 68 millisecond pulsars based on observations made between July 2004 and August 2020. Similar to \citet{nanograv_15yr_gwb}, we only analyze data from pulsars which have a timing baseline longer than three years. This results in the exclusion of PSR J0614$-$3329 and brings the total number of analyzed pulsars to 67. Compared to the analysis of the 12.5-year dataset, where the data from only 45 pulsars were analyzed \citep{nanograv_12p5yr_cw}, this represents an almost 50\% increase in the number of pulsars, and more than 20\% increase in timing baseline. In addition, we also benefit from improved timing solutions as the new dataset is a complete reanalysis of all previous data. A more detailed description of the dataset can be found in \citet{nanograv_15yr_dataset}.

\subsection{Models}
We model timing residuals in each pulsar as:
\begin{equation}
 \delta t = M \epsilon + n_{\rm WN} + n_{\rm RN} + n_{\rm CURN} + s,
\end{equation}
where $M$ is the design matrix that describes the linearized timing model and $\epsilon$ are the offsets from the nominal timing model parameters. The parameters $n_{\rm WN}$ and $n_{\rm RN}$ describe the white and red noise respectively in a particular pulsar, while $n_{\rm CURN}$ represents a common (spatially) uncorrelated red noise (CURN) signal present in all pulsars. More details on these noise models can be found in \citet{nanograv_15yr_detchar}. The GW signal of an individual binary (often called a continuous wave signal, or CW, due to its minimal frequency evolution) can be expressed as (see e.g.,~\citealt{nanograv_12p5yr_cw}):
\begin{eqnarray}
    s(t) &=& F^+(\theta, \phi, \psi) \; \left[s_+(t_p) - s_+(t)\right]\nonumber \\
    && + F^\times(\theta, \phi, \psi) \; \left[s_\times(t_p) - s_\times(t)\right] \,,
\end{eqnarray}
where $s_{+,\times}(t)$ and $s_{+,\times}(t_p)$ correspond to the so called Earth and pulsar terms, respectively; $t$ is the time measured at the Solar System barycenter; $t_p$ is the corresponding time at the pulsar, which depends on both the sky location and the distance of the pulsar; and $F^{+,\times}$ is the antenna pattern function, which depends on the sky location of the binary ($\theta$, $\phi$) and the GW polarization angle ($\psi$), and describes how the signal appears in a given pulsar (see eq.~(5) of \citealt{nanograv_12p5yr_cw}).

For a circular binary, at zeroth post-Newtonian (0-PN) order, $s_{+,\times}$ is given by:
\begin{eqnarray}
    s_+(t) &=& -\frac{{\cal M}^{5/3}}{d_L \, \omega(t)^{1/3}} \sin 2\Phi(t) \, \left(1+\cos^2 \iota \right) \,, \label{eq:signal1} \\
	s_\times(t) &=& \frac{{\cal M}^{5/3}}{d_L \, \omega(t)^{1/3}} 2 \cos 2\Phi(t) \, \cos \iota \,, \label{eq:signal2}
\end{eqnarray}
where $\iota$ is the inclination angle of the SMBHB, 
$d_L$ is the luminosity distance to the source, 
and ${\cal M} \equiv (m_1 m_2)^{3/5}/(m_1+m_2)^{1/5}$ 
is a combination of the black hole masses $m_1$ and $m_2$ 
called the ``chirp mass''.

The reference angular frequency of the Earth term is denoted as $\omega_0 = \omega(t_0) = 2 \pi f_{\rm GW}$. The time-dependent angular frequency is given by:
\begin{equation}
\omega(t) = \omega_0 \left[ 1 - \frac{256}{5} {\cal M}^{5/3}  \omega_0^{8/3} (t-t_0) \right]^{-3/8}.
\label{eq:omega_t}
\end{equation}
The frequency evolution within the observing time span is typically small. However, the pulsar terms are effectively sampling the waveform at earlier times depending on the pulsar distance and the relative position of the pulsar and the source on the sky. This means that the pulsar terms can have significantly lower frequencies, especially for high ${\cal M}$ and $f_{\rm GW}$ values. Note that both ${\cal M}$ and $\omega(t)$ are observer-frame quantities, which are related to their rest-frame equivalents as ${\cal M}_{\rm r} = {\cal M} / (1+z)$ and $\omega_{\rm r} = \omega (1+z)$.

The phases in eq.~(\ref{eq:signal1}) and (\ref{eq:signal2}) are given by:
\begin{eqnarray}
\Phi(t) &=& \Phi_0 + \frac{1}{32} {\cal M}^{-5/3} \left[ \omega_0^{-5/3} - \omega(t)^{-5/3} \right] \, ,
\end{eqnarray}
where $\Phi_0$ is the initial Earth term phase, and a similar expression holds for each pulsar phase with initial pulsar phases $\Phi_i$. We can also define the overall signal amplitude as:
\begin{equation}
    h_0 = \frac{2 \mathcal{M}^{5/3}(\pi f_{\rm GW})^{2/3}}{d_{L}}.
\label{eq:amplitude}
\end{equation}
Thus the signal model can be described with eight global parameters ($\theta$, $\phi$, $\psi$, $\iota$, $\Phi_0$, $f_{\rm GW}$, $\mathcal{M}$, $h_0$) and two additional parameters for each pulsar: the distance to the pulsar ($L_i$), and the phase of the signal at the pulsar ($\Phi_i$).

The white noise is modeled separately for each pulsar and backend-receiver combination and is described by three parameters (EFAC, EQUAD, ECORR), which we fix to their best-fit values found without modeling an individual binary, following previous searches (see e.g.,~\citealt{nanograv_12p5yr_cw}). We allow individual pulsar red noise to vary. An improvement over the previous search is that we also marginalize over the parameters of a CURN process.

\subsection{Sampler}
\label{ssec:sampler}
A key improvement over previous searches comes from using \texttt{QuickCW} \citep{QuickCW, QuickCW_code}, a software package that builds on the \texttt{enterprise} \citep{enterprise} and \texttt{enterprise\textunderscore extensions} \citep{ee} libraries, but uses a custom likelihood calculation and a Markov chain Monte Carlo (MCMC) sampler tailored to the search for individual binaries. The idea is that the expensive-to-calculate inner products needed for the likelihood can be stored and reused to calculate the likelihood at different values of the so-called projection parameters ($\psi$, $\iota$, $\Phi_0$, $h_0$, $\Phi_i$), when the rest of the parameters (shape parameters: $\theta$, $\phi$, $f_{\rm GW}$, $\mathcal{M}$, $L_i$) are held fixed. This results in an $\mathcal{O}(10^4)$ speed-up when updating projection parameters, which roughly translates to a 100-fold speed-up of the entire search. For more details see \citet{QuickCW}.

This increased efficiency also allows us to improve the search in several ways that were previously computationally prohibitive:
\begin{enumerate}
    \item Instead of fixing the amplitude and spectral index of the common red noise process as was done in \citet{nanograv_12p5yr_cw}, we can marginalize over those parameters. This is important due to the strong covariance between a low-frequency binary and the GWB.
    \item Instead of searching for binaries at a set of discrete frequencies as was done, e.g.,~in \citet{nanograv_12p5yr_cw}, we directly sample the GW frequency parameter in our MCMC. This makes our search fully Bayesian by removing a grid-based element of the previous search, and has the advantage of allowing the frequency parameter to take any value in the prior range.
    \item The improved efficiency makes it possible to repeat the analysis multiple times with various settings and to carry out more extensive tests of the algorithm (see Appendix \ref{sec:code_review}), both of which make the results more robust.
\end{enumerate}

While the frequencies are no longer tied to a grid, so in principle one could do a search over the whole prior range with a single MCMC run, in practice it can be beneficial to break the analysis up into frequency bands that are analyzed separately. The main reason for doing this is to ensure that all frequencies are well sampled; the sampler tends to favor some frequencies over others. In addition, it allows the analysis to be spread over the available computing resources. For the search (see Section \ref{sec:results}) we performed several independent MCMC runs either at low ($\leq25$ nHz) or high ($\geq25$ nHz) GW frequencies. This helped convergence substantially, since the model behaves differently in these two regimes. At low frequencies, there is a strong covariance with the CURN, but the CW signals evolve very little in frequency, so the pulsar distance parameters are practically irrelevant. At high frequencies, the red noise has no effect, but the signals can be strongly evolving, so the chirp mass and pulsar distance parameters become important. For the upper limit analysis (see Section \ref{ssec:UL}), we further subdivided the frequency range and performed analysis over nine overlapping frequency bins. This was necessary to get enough samples in each frequency bin, since the uniform amplitude prior used for these runs makes it harder to explore all frequencies compared to the uniform amplitude runs used to calculate Bayes factors.

\subsection{Priors}
We impose uninformative uniform priors on most model parameters to represent our lack of knowledge about them. For the common and individual pulsar red noise amplitudes, we use a log-uniform prior, which allows the sampler to explore amplitudes over many orders of magnitude. For the amplitude of the individual binary signal, we use a log-uniform prior when calculating Bayes factors, and a uniform prior for calculating upper limits.

The pulsar distances are the only parameters where we do have prior knowledge. For these we adopt the approach introduced in \citet{nanograv_12p5yr_cw} of using the best available distance measurements to set up our priors for each pulsar. The form of these priors depend on the source of the distance measurement. For pulsars with a precise parallax measurement,\footnote{Based on the list of pulsar parallax measurements maintained at: \url{http://hosting.astro.cornell.edu/~shami/parallax/}} we assume a Gaussian distribution on the parallax, which results in a skewed prior distribution of the distance (see eq.~(20) in \citealt{nanograv_12p5yr_cw}). For other pulsars we converted the measured dispersion measure value to a distance estimate using the NE2001 electron density map \citep{ne2001}. Following \citet{nanograv_12p5yr_cw} we assumed a 20\% error on these estimates and set the prior distribution to be uniform between 80\% and 120\% of the measured value with Gaussian tails on either side with standard deviation of 5\% of the measured value (see eq.~(21) in \citealt{nanograv_12p5yr_cw}). We list the parameters we used for each pulsar in Table \ref{tab:psr_dists} in Appendix \ref{sec:psr_dist}.

{Note that these are the same priors as used in the NANOGrav 12.5-year individual binary search (see Table 1 in \citealt{nanograv_12p5yr_cw}), except that: i) we use a log-uniform prior on the GW frequency instead of fixing its value on a grid; ii) we vary the parameters of the common red noise instead of fixing them; and iii) we updated our pulsar distances to include recent parallax measurements (see Table \ref{tab:psr_dists}).

\subsection{Accounting for HD Correlations with Resampling}

\begin{figure*}[htbp]
    \centering
    \includegraphics[width = 2\columnwidth]{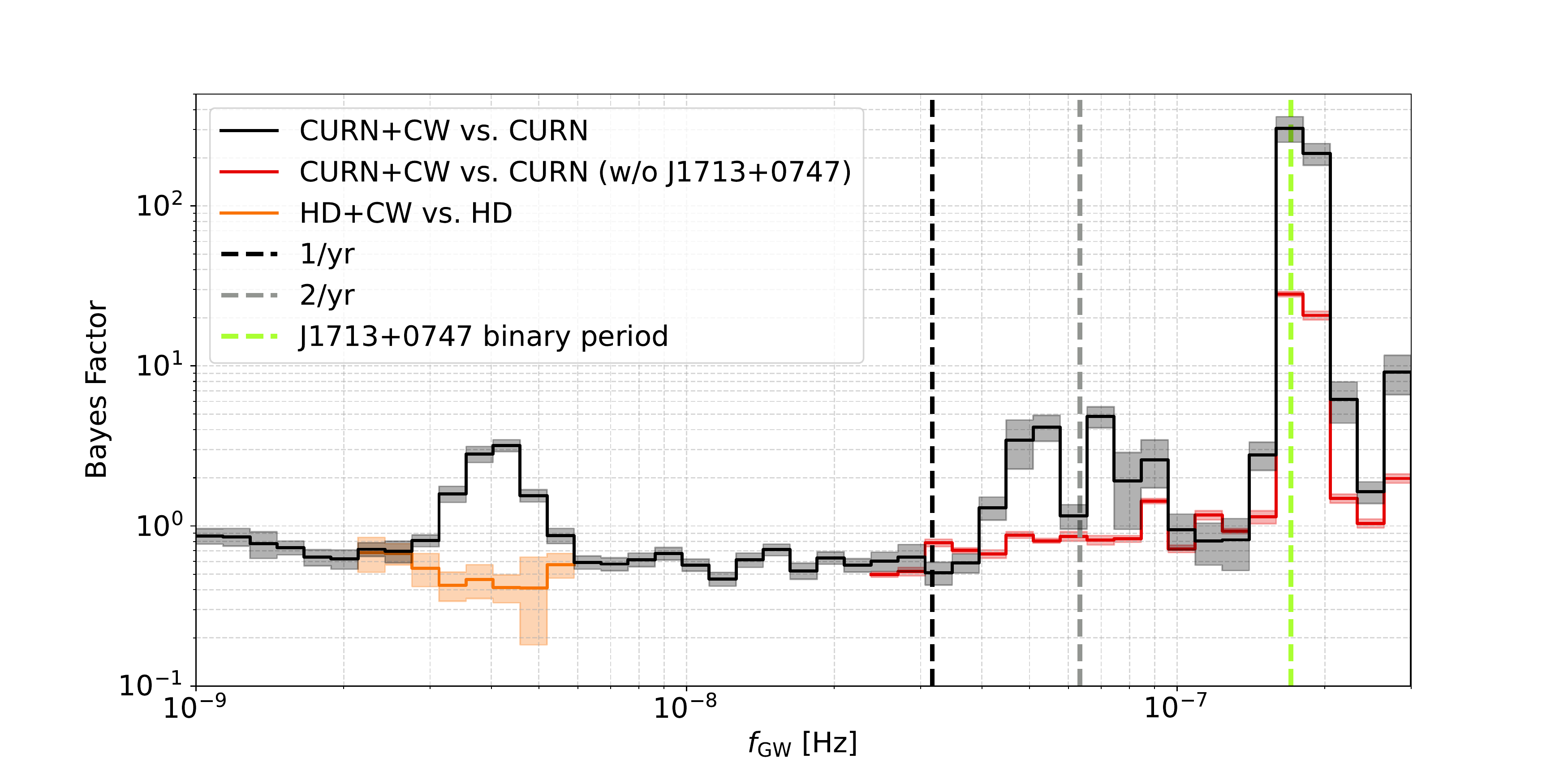}
    \caption{Savage-Dickey Bayes factors for the CW+CURN model versus the CURN model as a function of frequency (black). Also shown are Bayes factors when excluding PSR J1713+0747 (red, only computed for $f_{\rm GW}>24$ nHz) and Bayes factors based on a resampled posterior that takes into account the presence of HD correlations in the common red noise process, i.e., CW+HD versus HD (orange, only computed for 2.1 nHz $<f_{\rm GW}<5.9$ nHz). Shaded regions show the 1-$\sigma$ uncertainties.}
    \label{fig:bf}
\end{figure*}

\label{ssec:reweighting}
The common red noise process was shown to have correlations between pulsars that follow HD correlations \citep{nanograv_15yr_gwb}, so for a fully consistent modeling, we should include those correlations when searching for an individual binary. However, it is currently computationally prohibitive to directly model those correlations while also searching for signals from individual binaries. Instead, we model the background as uncorrelated common red noise and take the correlations into account in post-processing.

We do so by using the likelihood reweighing technique introduced to PTA data analysis by \citet{sophie_resampling}. The idea is to take a thinned set of posterior samples assuming a CURN, and calculate importance weights based on the ratio of the likelihood with (HD) and without correlations (CURN). These weights can be used with importance sampling to produce correct posterior samples with correlations taken into account. As a result, we have four different models to explore, which are summarized in Table \ref{tab:modeltab}. In addition, these weights can also be used to calculate a Bayes factor between the HD and CURN models.

\begin{table}[htb]
\begin{center}
\caption{Data models.}
\label{tab:modeltab}
\centering
\begin{scriptsize}
\begin{tabular}{l|cccc}
\hline \hline
 & CURN & HD & CURN+CW & HD+CW \\
\hline
 Individual binary &  &  & \cc & \cc \\
\hline
 Hellings-Downs &  & \multirow{2}{*}{\cc} &  & \multirow{2}{*}{\cc} \\
 spatial correlations & & & & \\
\hline 
Common red noise & \cc & \cc & \cc & \cc \\
 \hline
 Pulsar-intrinsic		& \multirow{2}{*}{\cc} & \multirow{2}{*}{\cc} & \multirow{2}{*}{\cc} & \multirow{2}{*}{\cc} \\ 
 red noise	 &	   &	 &     &    \\
\hline \hline
\end{tabular}
\end{scriptsize}
\end{center}
\end{table}

\section{Search Results}
\label{sec:results}

We performed an all-sky search for individual SMBHBs with GW frequencies between 1 and 300 nHz, roughly corresponding to $0.5/T_{\rm obs}$\footnote{Note that while PTAs lose sensitivity at frequencies below $1/T_{\rm obs}$ due to the fit to pulsar spindown, the sensitivity does not diminish completely (see e.g.,~\citealt{jeff_sensitivity}), allowing us to search for signals below $1/T_{\rm obs}$.} and $150/T_{\rm obs}$, where $T_{\rm obs}$ is the total timespan of the dataset. We used a log-uniform prior both on the GW amplitude and the GW frequency. We calculate the Bayes factor in favor of including a binary using the Savage-Dickey density ratio \citep{SD_BF}. This relies on the fact that the model with the binary reduces to the noise-only model when the amplitude is zero, and the Bayes factor is given by the density ratio of the posterior and the prior at zero amplitude. We approximate the posterior density by binning $\log_{10} h_0$ samples and taking the density at the lowest amplitude bin. The statistical error of the Bayes factor is estimated from the standard deviation of the ten lowest amplitude bins, since those all correspond to practically zero amplitude.

Figure \ref{fig:bf} shows the Bayes factors in favor of a binary as a function of $f_{\rm GW}$ (black curve) along with its 1-$\sigma$ uncertainties (gray shaded region). We show the results over 37 log-uniform frequency bins. Over most of the frequency bins, the presence of a binary is either disfavored (Bayes factor$<$1) or the analysis is indecisive (Bayes factor$\sim$1). However, there are a few frequencies, where the Bayes factor indicates some support for a signal. One distinct peak is visible around 4 nHz ($\sim$2/$T_{\rm obs}$), and there are several peaks at high frequencies, most notably around 170 nHz. Below we discuss these candidates in more detail.

\subsection{Low-frequency Candidate}
\label{ssec:lowf_candidate}

\begin{figure*}[htbp!]
    \centering
    \includegraphics[width = 2\columnwidth]{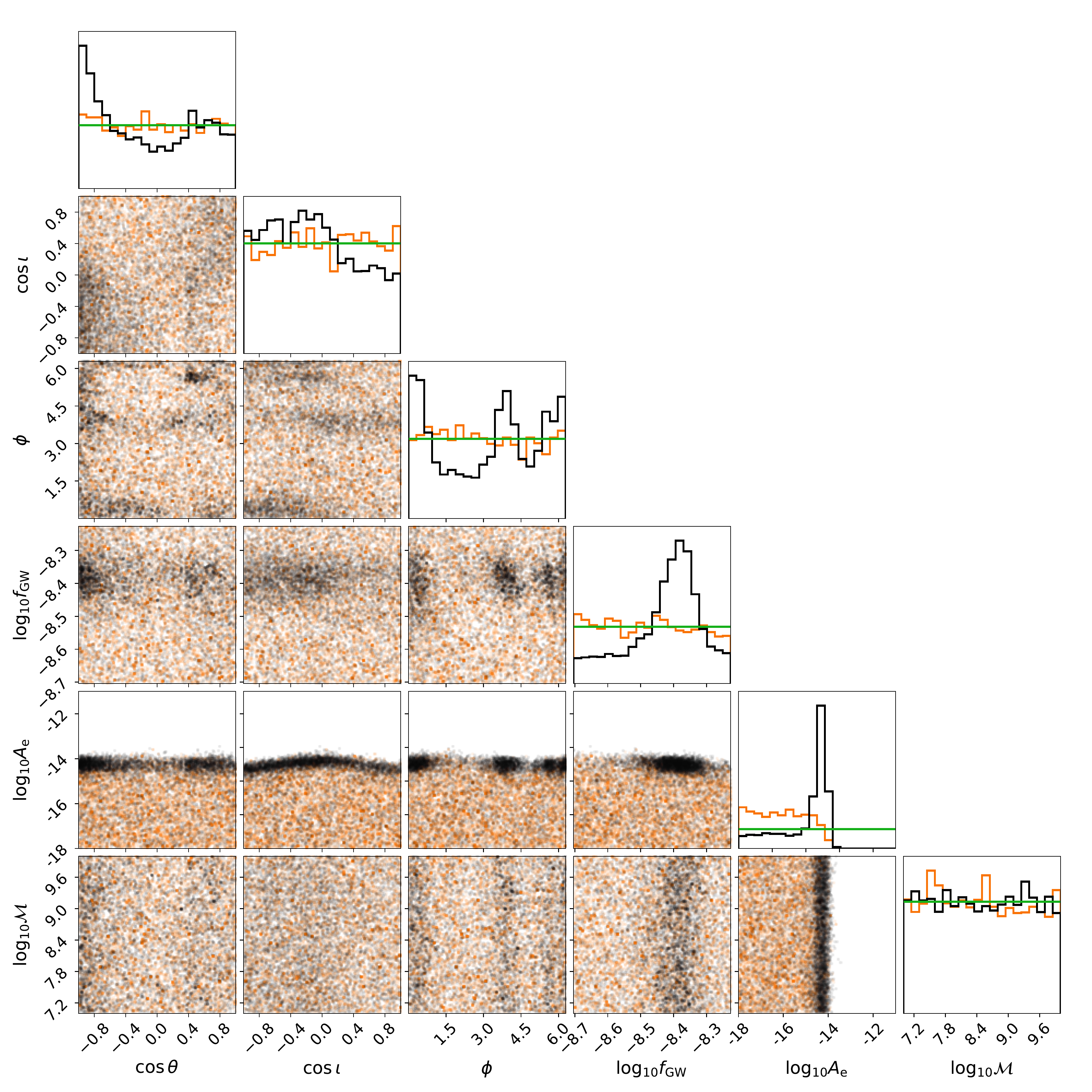}
    \caption{Distribution of model parameters for the low-frequency candidate at 4 nHz before (black) and after (orange) resampling to take HD correlations into account. Green lines show the prior distributions for each parameter. Note that after resampling, the peak in the amplitude posterior disappears and all other parameters follow their prior distributions, thus no longer showing support for the candidate signal.}
    \label{fig:lowf_candidate_corner}
\end{figure*}

We find that the presence of an individual binary with $f_{\rm GW}\sim 4$ nHz is mildly favored over only having a CURN process (Bayes factor$\sim$3). However, since \citet{nanograv_15yr_gwb} found evidence for the red noise process having HD-correlations (HD versus CURN Bayes factor $\sim200$), the right models to compare are an HD-correlated background vs an HD-correlated background and an individual binary. We used methods outlined in Section \ref{ssec:reweighting} to take the correlations into account by reweighing the posterior samples from an analysis including uncorrelated red noise. The resulting Bayes factors for the presence of an individual binary are shown in Figure \ref{fig:bf} with the orange curve. Including the HD correlations leads to the low-frequency CW candidate being disfavored, with the HD-only model mildly preferred over the HD+CW model by a Bayes factor of $\sim$2.5.

Figure \ref{fig:lowf_candidate_corner} shows the distribution of a few key binary parameters before (black) and after (orange) taking HD-correlations into account. Green horizontal lines indicate the prior distribution for each parameter. The results before reweighing have an amplitude posterior with a strong peak at a non-zero value, and all other parameters show informative distributions. However, after reweighing, the amplitude posterior prefers low values and all other parameters return their respective prior distributions. Including the correlations removes any sign of the putative low-frequency signal. One caveat of this reweighing technique is that we cannot be sure that the true HD+CW and CURN+CW posteriors have sufficiently similar posteriors for the reweighing to work effectively. It is possible that while a signal with the parameters found by the CURN+CW search is disfavored in the HD+CW model, there might be a CW signal with different parameters favored in the HD+CW model. One indication of this limitation is that the reweighed samples recover the prior, while for a full HD+CW search we would expect to see some structure in e.g.,~the sky location due to the highly anisotropic sensitivity of the PTA.

Interestingly, the frequency of this low-frequency candidate is consistent with the subdominant monopolar signature that was found along the HD process in \citet{nanograv_15yr_gwb}. Thus we investigated if an individual binary could account for that monopolar signature by calculating signal-to-noise ratio (S/N) of the noise-marginalized multi-component optimal statistic (MCOS, \citealt{os,nmos,mcos}), which is a frequentist detection statistic for the presence of processes with different spatial correlation patterns (monopole, dipole, HD). We do so under two model assumptions: i) the dataset is solely described by a common correlated red noise process (CCRN) with correlations modeled by the MCOS; ii) the dataset is descibed by an individual binary and a CCRN (CCRN+CW), so the MCOS only needs to fit correlations that remain after modeling out the deterministic CW signal. In practice the noise-marginalized MCOS gives a distribution of S/N values by calculating them at random draws from a Bayesian CURN (or CURN+CW) posterior.\footnote{Thus the parameters of the CW signal removed are also marginalized over, instead of simply removing a point estimate.} The top panel of Figure \ref{fig:OS} shows the distribution of the noise-marginalized MCOS S/N for monopolar and HD correlations under the CCRN model (dipolar correlations were also modeled, but they consistently return S/N$\sim$0), while the bottom panel is the same but under the CCRN+CW model. We also show the means of these distributions (dashed lines) and the maximum-likelihood S/N values (solid lines). For the CCRN model, the HD process has a higher mean and maximum-likelihood S/N, but the distribution overlaps significantly with the distribution for the monopole S/N. We can see that once the CW signal is included in the model, the significance of both processes are reduced, with the distributions allowing for S/N=0, and maximum-likelihood S/N values below 1 for both processes.

\begin{figure}
    \centering
    \includegraphics[width = 1\columnwidth]{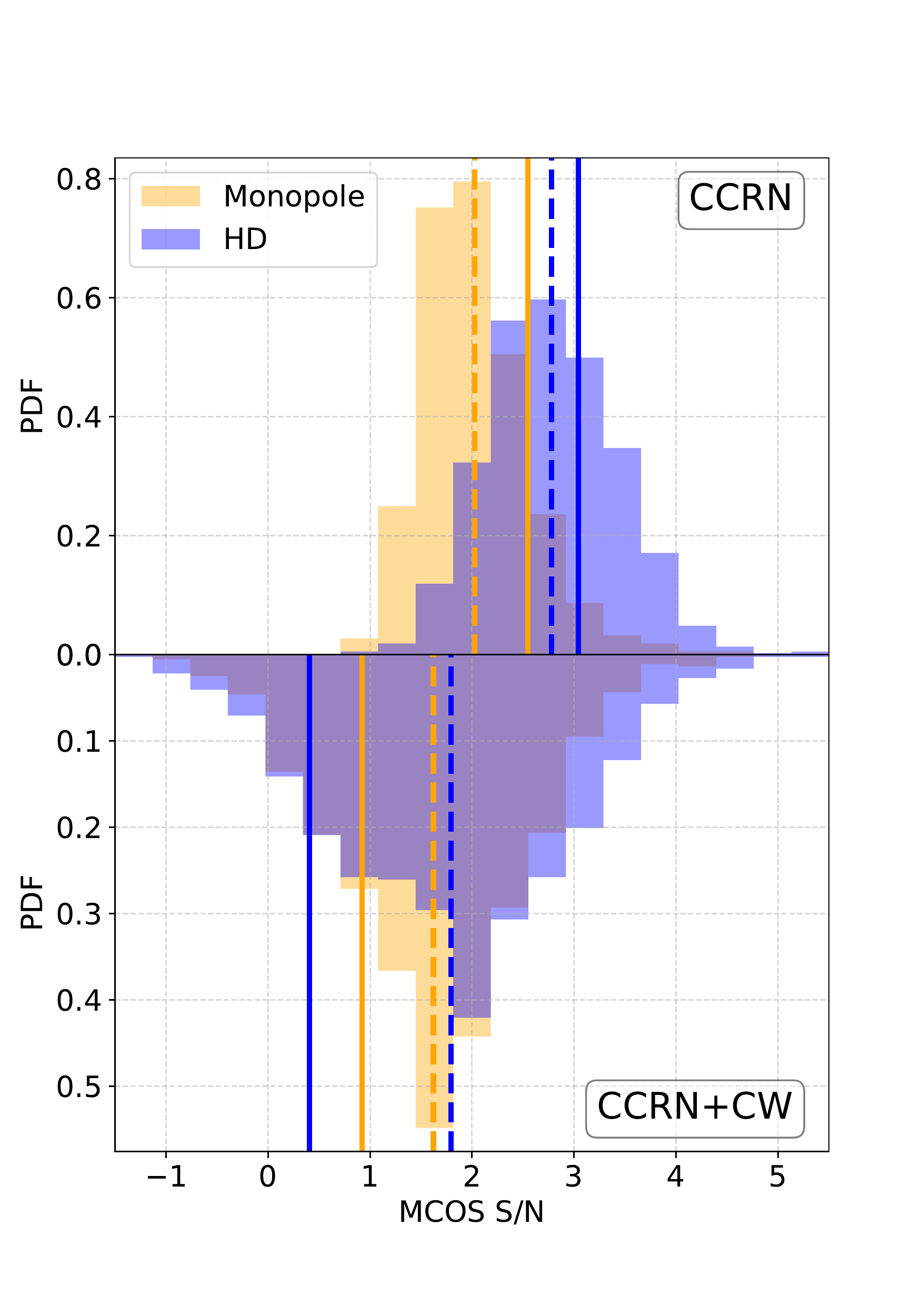}
    \caption{Noise-marginalized MCOS S/N distributions of different spatially correlated processes with (bottom panel) and without (top panel) the binary signal removed. Solid vertical lines show the maximum likelihood values, while dashed lines indicate the mean of each distribution. Removing the binary candidate around 4 nHz is able to shift the distribution for both processes so that they become consistent with zero.}
    \label{fig:OS}
\end{figure}

Since even a single CW signal induces HD correlations in pulsar timing array data \citep{Cornish:2013aba}, it is not surprising that removing the candidate CW signal reduces the S/N of the HD process. In the limit of an array having a large number of isotropically distributed pulsars, a single CW induces pure averaged HD correlations, but for a finite array, with an anisotropic pulsar distribution, some amount of monopole and dipole correlations are also produced.\footnote{An anisotropic pulsar array responding to a single CW is mathematically equivalent to an isotropic array responding to an anisotropic stochastic signal (this follow from equations A21 and A11 of \citet{Allen:2022dzg}, and replacing the angular average in A15 with an average weighted by the spherical harmonic expansions of the pulsar distribution). For example, as shown in \citet{Mingarelli:2013dsa}, a dipole anisotropy results in a mix of HD, monople, dipole and quadrupole corrleations.} This explains why the monopole S/N is also changed when the CW model is removed.

We summarize the Bayes factors between each model combination in Figure \ref{fig:bf_graph}. The HD vs.~CURN Bayes factor is based on a hyper-model run \citep{nanograv_15yr_gwb}, the CURN+CW vs.~CURN and HD+CW vs.~HD Bayes factors are calculated via the Savage-Dickey density ratio for the GW frequency bin between 4 and 4.6 nHz (c.f.~Figure \ref{fig:bf}), and the HD+CW vs.~CURN+CW Bayes factor is calculated through the reweighing process \citep{sophie_resampling}. The HD+CW vs.~CURN Bayes factor can be calculated from these Bayes factors in two different ways, either as a product of the CURN+CW vs.~CURN and HD+CW vs.~CURN+CW or of the HD vs.~CURN and HD+CW vs.~HD Bayes factors. The two values are indicated on the diagonal on Figure \ref{fig:bf_graph}, and serve as a consistency check of the results. Note that while the inclusion of a CW signal is disfavored if we restrict the background to be HD-correlated (Bayes factor of 0.4), it is also true that the evidence for HD correlations is significantly reduced if we require a CW signal in the model (Bayes factor changes from 190 to 35). This indicates that the HD and CW models have some covariance, however, overall the most preferred model among these four is HD. 

\begin{figure}
    \centering
    \includegraphics[width = 1\columnwidth]{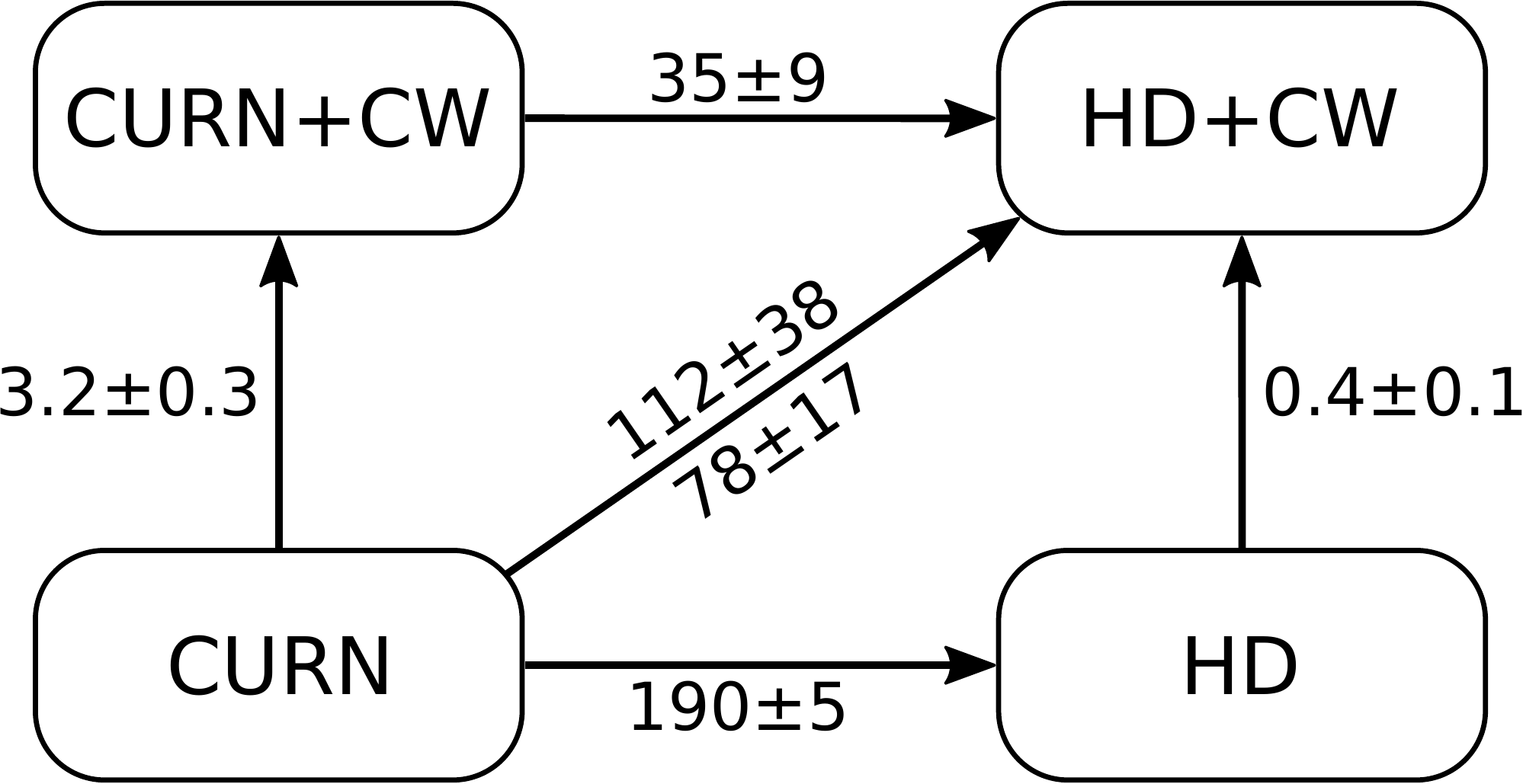}
    \caption{Bayes factors between different combinations of CURN, HD, and CW models. CW here refers to an individual binary with frequencies between 4 nHz and 4.6 nHz, the frequency bin where the significance of the low-frequency candidate peaks (c.f.~Figure \ref{fig:bf}). The two different values on the diagonal correspond to two different ways of calculating that Bayes factor (going through the CURN+CW or the HD model) and serves as a consistency check.}
    \label{fig:bf_graph}
\end{figure}

To summarize, we find (1) mild evidence for the CW+CURN model versus the CURN model; (2) mild evidence for the HD model over the HD+CW model; and (3) reduction in the S/N for the HD model when the CW model is subtracted. One interpretation of these results is that the HD and CW models are competing over the same signal, and at this juncture, the simpler HD model provides the better description in a Bayesian statistical sense. It will be interesting to see if the evidence shifts as more data are accumulated, especially given that 4 nHz is right around the frequency where population models suggest we should make the first CW detection \citep{Luke_single_source, Becsy:2022pnr}.

\subsection{High-frequency Candidate}
\label{ssec:highf_candidate}

We find the highest support for an individual binary at frequencies between 159 nHz and 181 nHz, where the Bayes factor in favor of the binary is $306\pm56$ (see Figure \ref{fig:bf}). There is also strong support in the next frequency bin, between 181 nHz and 205 nHz, where we find a Bayes factor of $212\pm33$. These frequencies are close to the binary period of one of the longest timed NANOGrav pulsars, PSR J1713+0747 (170.6 nHz). To investigate if there could be an interaction between the GW and the timing models, we reconstructed the timing model perturbations in PSR J1713+0747 using methods described in \citet{meyers+23}. We find that the amplitude of the timing model perturbations and the CW signal at this frequency are similar, suggesting that there can be an interplay between the two models. This might be alleviated by sampling the timing model parameters during the search, instead of using the linearized timing model. We do not currently have the infrastructure to do so, but this could be an important direction to explore in the future (for a discussion of full Bayesian timing see \citealt{bayesian_timing}).

Instead, to assess the contribution of PSR J1713+0747 to this candidate, we reran the analysis on all the pulsars except PSR J1713+0747. The resulting Bayes factors are shown by the red line in Figure \ref{fig:bf}. This change removed many moderate Bayes factor peaks at high frequencies, and also reduced the significance of the largest peak from $\sim$300 to $\sim$30. However, since this is still the most significant candidate, even after removing PSR J1713+0747, we cannot conclude that the high Bayes factor is solely due to the interplay with PSR J1713+0747's binary model. In addition, we also find that the Bayes factor can vary significantly depending on the noise models used, suggesting covariance with unmodeled high-frequency noise (see Appendix \ref{sec:adv_noise} for more details).

It is important to note that our prior expectation is that it is very unlikely to have a detectable binary at such high frequencies \citep{SVV2009, rosado_expected_properties, Luke_single_source, Becsy:2022pnr}, mainly because of their relatively faster evolution (and thus smaller residence time) compared to binaries at lower frequencies. In addition, we compared the reconstructed sky location and distance of the source with the NANOGrav galaxy catalog \citep{Arzoumanian+2021} which contains 45,000 galaxies within 500 Mpc, enhanced with important quantities for GW detection (e.g., SMBH mass and galaxy distance). This catalog is complete to a K-band magnitude of 11.75, which means that for the small inferred distance ($\lesssim$ 20 Mpc), it is complete to galaxies with an SMBH mass above $\sim 3\times10^6 \ M_{\odot}$. We only find 8 galaxies in the localization volume, the largest of which has a predicted largest possible chirp mass of $2\times10^7 \ M_{\odot}$, which is an order of magnitude smaller than the chirp mass we recover for this candidate ($\gtrsim 2\times10^8 \ M_{\odot}$).

Given the high likelihood of unmodeled high-frequency noise, the interplay with the timing model, and the implausible astrophysical parameters of the candidate, we conclude that it is unlikely that this candidate signal is due to a SMBHB. 

\section{Limits}
\label{sec:limits}

\subsection{Amplitude Upper Limits}
\label{ssec:UL}
As we find no compelling evidence for individual binaries, we place upper limits on their strain amplitudes. These should be calculated with a uniform amplitude prior, as with log-uniform priors the upper limit strongly depends on the lower prior bound. See Appendix \ref{sec:UL_prior} for more discussion on the effect of the prior choice and HD correlations.

Figure \ref{fig:ul_progress} shows our upper limit using uniform amplitude priors, along with upper limits from the NANOGrav 11-year \citep{nanograv_11yr_cw} and 12.5-year datasets \citep{nanograv_12p5yr_cw}. The 15-year dataset produces improved upper limits at all frequencies below $\sim$6 nHz, where the increased timespan and the fact that we allow the GWB parameters to vary matter the most. Above this frequency the 15-year upper limits are roughly similar to previous ones, with improved upper limits at some frequencies and weakened upper limits at others. Note however that a direct comparison with earlier results is not possible due to the differences in methodology, most notably due to the fact that we search over the GW frequency and bin the samples in postprocessing, while earlier results were produced with individual runs at fixed frequencies. In addition, our model is also different, since we marginalize over the CURN instead of fixing its parameters. However, if the upper limits are indeed experiencing a stagnation, a possible explanation of that could be unmodeled noise at high frequencies, the effect of which could be reduced in the future by the full adoption of advanced noise models \citep{adv_noise}. In Appendix \ref{sec:UL_12p5yr} we further investigate the comparison with the 12.5-year results by reanalyzing the NANOGrav 12.5-year dataset with the methods used in this search.

\begin{figure}[h]
    \centering
    \includegraphics[width = 1\columnwidth]{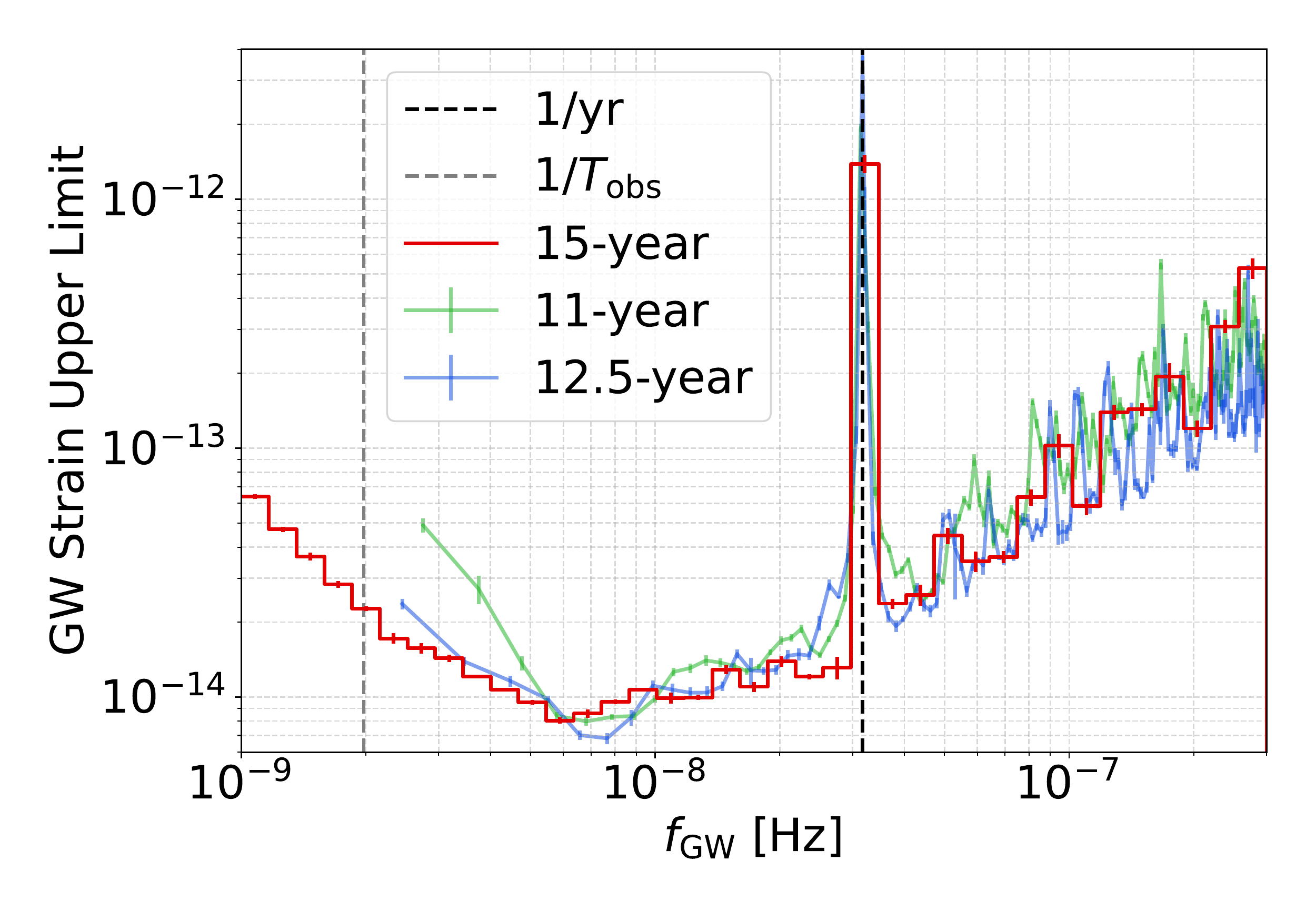}
    \caption{All-sky CW strain 95\% upper limits compared to previous datasets. We see the most significant improvements at low frequencies where marginalizing over the common red noise allows us to set more constraining upper limits.}
    \label{fig:ul_progress}
\end{figure}

In Figure \ref{fig:map_6nHz} we show the strain upper limit on the sky at the most sensitive frequency of 6 nHz. Red stars indicate the location of the 67 pulsars used for the analysis. Note that the anisotropic distribution of pulsars results in a highly anisotropic sensitivity on the sky, with about a factor of 7 difference between the most and least sensitive sky location. We also indicate the sky-averaged upper limit on the color scale (white vertical line). Note that the sky-averaging is not done uniformly on the sky, but rather through the posterior samples, which in practice results in the all-sky limit being biased high. In this example $\sim$73\% of the sky gives a lower upper limit than the all-sky value. The white cross marks the sky location with the lowest strain upper limit.

\begin{figure}
    \centering
    \includegraphics[width = 1\columnwidth]{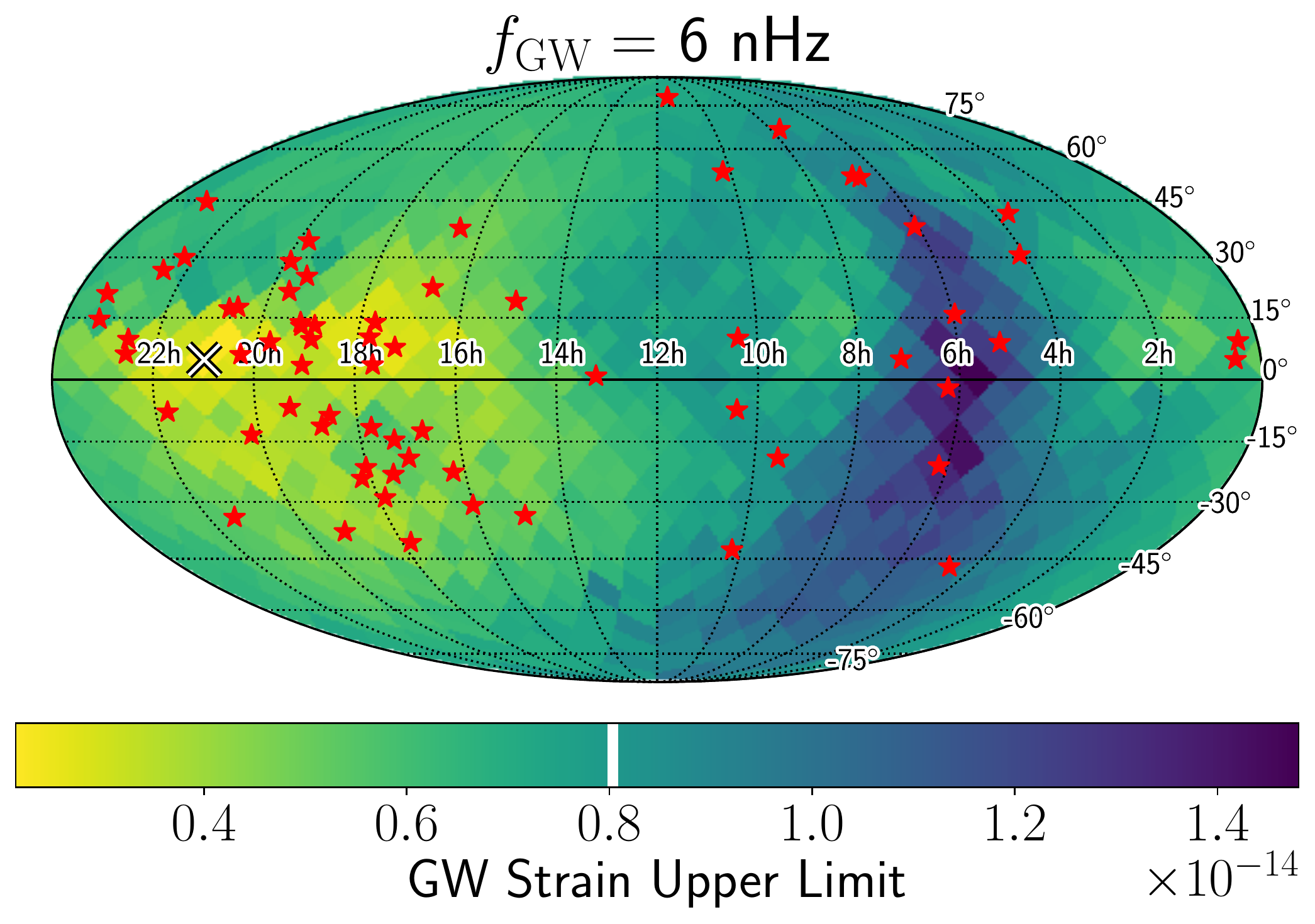}
    \caption{Map of CW strain 95\% upper limits at 6 nHz, the most sensitive frequency (see Figure \ref{fig:ul_progress}). Red stars indicate the location of pulsars included in the dataset. Note that the upper limit is almost an order of magnitude lower towards the galactic center towards which most of the pulsars are located. The white line on the color scale indicates the upper limit we find when marginalizing over sky location.}
    \label{fig:map_6nHz}
\end{figure}

Figures~\ref{fig:ul_progress} and \ref{fig:map_6nHz} indicate that our sensitivity depends strongly on both frequency and sky location. While we expect that the overall dipolar structure will be present in the sky location dependence of the sensitivity at all frequencies, there is no reason to expect the upper limit skymap to show the same structure at all frequencies. Thus, to allow follow-up studies to use the full search results, we release the upper limits as a function of frequency and sky location as supplemental materials \citep{nanograv_15yr_cw_sup_mat}.

\subsection{Distance Limits}
\label{sec:dist}

The amplitude of a CW signal depends on the GW frequency, the chirp mass, and the luminosity distance, as indicated in eq.~(\ref{eq:amplitude}). Thus we can also place constraints on the luminosity distance if we assume a chirp mass value.\footnote{In principle the chirp mass could be constrained by the analysis, but in practice it is entirely unconstrained at low frequencies and only weakly constrained at high frequencies. Thus using the chirp mass values from the analysis would practically be the same as fixing the chirp mass to its lower prior bound.} Given a fiducial chirp mass value, we can turn the all-sky strain upper limit into an all-sky luminosity distance lower limit in every frequency bin. We show this for ${\cal M}=10^9 \ M_{\odot}$ in Figure \ref{fig:dist} (red line), which can be easily scaled to any chirp mass value. However, as we have seen above, such a limit is dominated by the least sensitive area of the sky. An alternative way of calculating a distance limit that alleviates that bias is as follows. We can calculate the distance upper limit at each frequency and at each pixel on the sky.\footnote{Here we use a low number of pixels (12) to make sure that all pixels at all frequencies have sufficient number of samples.} This gives us a 3-dimensional exclusion volume ($V_{\rm ex}$) at each frequency, which is highly asymmetric due to the anisotropic sensitivity of the array. We can define an effective radius, $R_{\rm eff}=[3V_{\rm ex}/(4\pi)]^{1/3}$, such that the volume of a sphere with that radius is the exclusion volume we find at that frequency. The blue line in Figure \ref{fig:dist} shows $R_{\rm eff}$ as a function of frequency. As a reference, we also show the range of distance limits between the best and worst sky pixel (blue shaded region). We can see that the all-sky limit and $R_{\rm eff}$ are always between the worst and best sky location limits, as they represent two different ways of averaging over the sky. Note also that $R_{\rm eff}$ is always larger than the all-sky limit, since it is less biased by the worst sky location. In addition, $R_{\rm eff}$ is easier to interpret due to its connection with the volume in which we can rule out the existence of SMBHBs.
 
\begin{figure}
    \centering
    \includegraphics[width = 1\columnwidth]{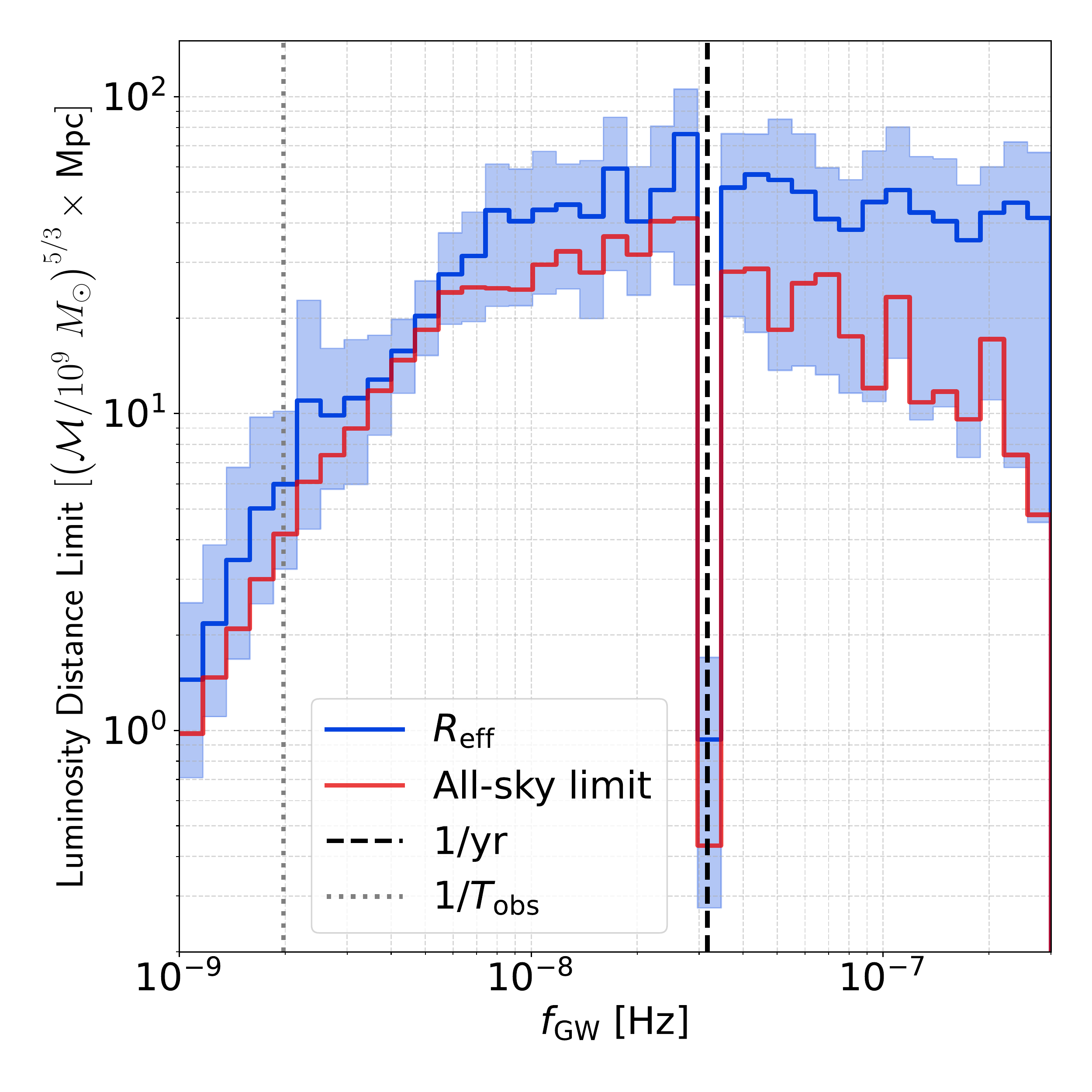}
    \caption{The 95\% lower limits on the luminosity distance to an individual SMBHB. The blue line shows the effective luminosity distance radius corresponding to the exclusion volume. The blue shaded region shows the range of distance limits over 12 equal-area sky pixels. The red line shows the all-sky distance limit calculated from the all-sky strain upper limit. Note that this is dominated by the worst sky location at every frequency.}
    \label{fig:dist}
\end{figure}

\begin{figure*}
    \centering
    \includegraphics[width = 1.5\columnwidth]{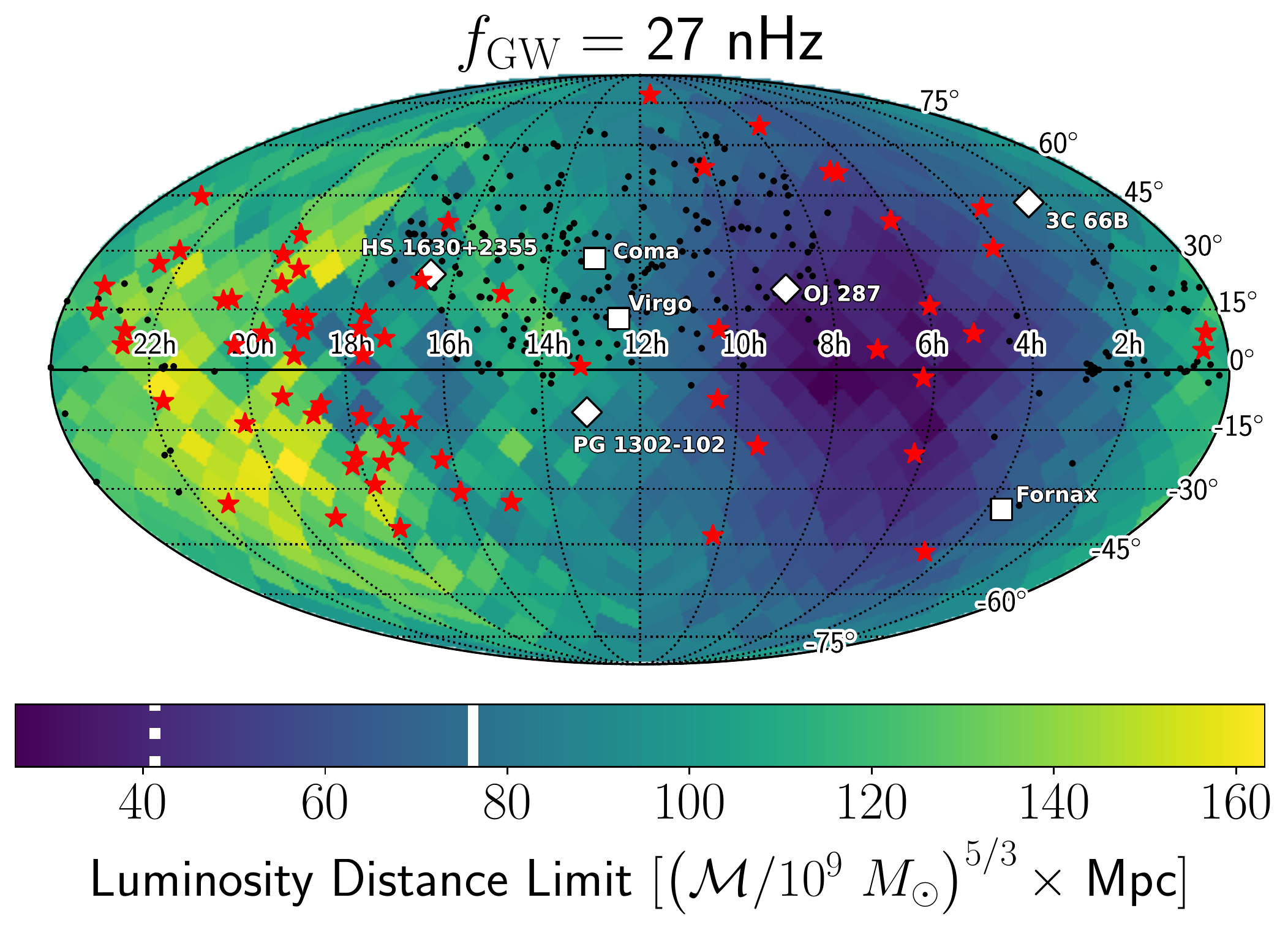}
    \caption{Map of the 95\% lower limit on the distance to individual SMBHBs with $\mathcal{M} = 10^9 M_\odot$ and $f_{\rm GW} = 27$ nHz. White diamonds indicate the positions of known SMBHB candidates, and white squares show the location of large galaxy clusters that could contain an SMBHB. Black dots mark the location of quasars with periodic variability. We show the all-sky limit (dotted white) and $R_{\rm eff}$ (solid white) on the color scale.}
    \label{fig:dist_map}
\end{figure*}

Figure \ref{fig:dist} can be used as a first step in evaluating SMBHB candidates found through electromagnetic observations. If a given candidate lies below the shaded region, we should have seen it in the search regardless of its sky location, if it lies above the shaded region, we would not be able to see regardless of its location on the sky. If a candidate lies within the blue shaded region, then its detectability depends on its sky location. To allow for further tests like that, we release our full 3-dimensional luminosity distance limits as a function of frequency and sky location as supplemental materials \citep{nanograv_15yr_cw_sup_mat}. As an example, Figure \ref{fig:dist_map} shows the luminosity distance limit as a function of sky location at the frequency where the search is sensitive to the largest distance (27 nHz). Note that this is different than the frequency where we can set the lowest strain amplitude upper limit due to the frequency-dependent relation between distance and amplitude, see eq.~(\ref{eq:amplitude}). We also show with white diamonds some notable SMBHB candidates, including 3C66B, for which PTAs provided the first multi-messenger constraints \citep{Jenet+2004, caitlin_3c66b}, OJ287 a bright periodic blazar candidate with a lightcurve that spans over a century \citep{Komossa+2023, Valtonen+2023}, PG1302-102 the first and most well-studied of the periodic quasar candidates from the modern time-domain surveys \citep{Graham+2015} and HS1630+2355, which is ranked highest based on its probability for detection by PTAs \citep{Xin_Mingarelli+}. Additional candidates detected as quasars with periodic variability from recent systematic searches in time-domain surveys \citep{Graham+2015b,Charisi+2016,LiuGezari+2019,Chen_DES_PLCs+2020,Chen_ZTF_PLCs+2022} are shown with black dots. A few close-by galaxy clusters (Coma, Virgo, Fornax) are also shown with white squares. 

\subsection{SMBHB Number Density Limits}

The exclusion volume defined above can also be used to set upper limits on the number density of SMBHBs with given chirp mass and GW frequency. We adopt the approach introduced in \citet{nanograv_12p5yr_cw} to convert an exclusion volume into a number density limit at some confidence level ($p_0$) by assuming these sources follow a Poisson distribution and integrating it out to get the desired confidence level, resulting in a number density upper limit of:
\begin{equation}
n_{\rm UL} = \frac{-\ln (1-p_0)}{V_c},
\end{equation}
where $V_c$ is the excluded volume measured in comoving distance. We convert luminosity distance limits to comoving distance limits assuming cosmological parameters from \citet{wmap9} using the \texttt{cosmopy} package \citep{cosmopy}.

Figure \ref{fig:smbhb_number_density} shows the number density upper limits we get with a confidence level of $p_0=0.95$. We show these results for five different fiducial chirp mass values. As expected we can put more stringent constraints on high-mass systems. Unlike the luminosity distance limits in Figure \ref{fig:dist}, these can only be approximated at any desired chirp mass with the $\mathcal{M}^{5/3}$ scaling, because the conversion between luminosity distance and comoving distance is nonlinear.

\begin{figure}
    \centering
    \includegraphics[width = 1\columnwidth]{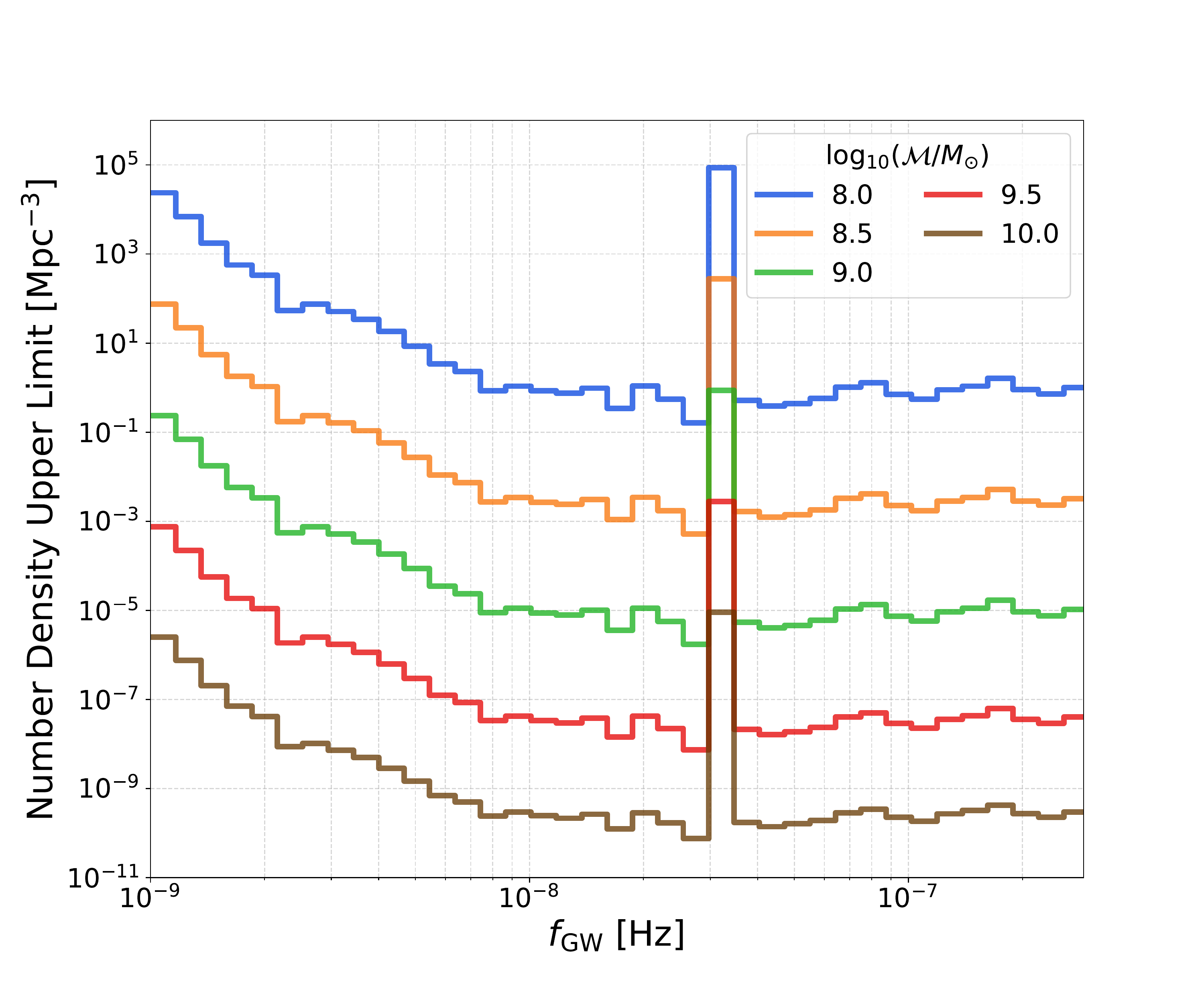}
    \caption{Upper limit on the number density of SMBHBs in the local universe as a function of GW frequency. Different curves correspond to binaries with different chirp mass values. Note that as expected, we can place more stringent limits on the number density of systems with higher masses. The density is measured in terms of comoving volume.}
    \label{fig:smbhb_number_density}
\end{figure}

\section{Conclusions}
\label{sec:conclusion}
We used efficient Bayesian methods to search for individual SMBHBs in the NANOGrav 15-year dataset \citep{nanograv_15yr_dataset}. This is the first time the \texttt{QuickCW} algorithm \citep{QuickCW} was employed on real data, which significantly reduced the computational burden and resulting environmental impact of this analysis. Using \texttt{QuickCW} also allowed us to improve the analysis in several ways. We were able to search over the frequency of the individual binary instead of carrying out a grid search as was done in previous analyses (see e.g.,~\citealt{nanograv_12p5yr_cw}). We also searched for the first time for an individual binary simultaneously with a common uncorrelated red noise process. 

We found a few frequencies where the presence of an individual binary was slightly favored over the noise-only model. Most of these candidates are at high frequencies, where the likely presence of unmodeled noise and our astrophysical prior expectations make these unlikely to be due to individual SMBHBs. The high-frequency candidate with the highest significance also has a localization volume with no plausible host galaxies, thus further lowering the plausibility of it being an SMBHB. We also found a candidate at $\sim$4 nHz, which is more in line with our expectations of where individual binaries are likely to be first detected \citep{Luke_single_source,Becsy:2022pnr}. However, the model including an individual binary at this frequency is only marginally favored (Bayes factor $\sim$3), and it becomes disfavored (Bayes factor $\sim$0.4) once we take into account the correlations present in the GWB. 

We set updated upper limits on the strain amplitude of GWs from individual binaries. These show improvement over previous datasets at low frequencies ($\lesssim$ 6 nHz) where the longer dataset timespan and marginalizing over the common red noise makes a difference. We also developed a new way of accounting for the highly anisotropic sensitivity of our array when quoting limits on the distance and number density of SMBHBs.

With recent results reporting evidence for the presence of a GWB in PTA datasets \citep{nanograv_15yr_gwb, epta_dr2_gwb, ppta_dr3_gwb}, searches for individual binaries are becoming more and more important. Assuming the GWB is due to a population of SMBHBs, we can reasonably expect to also see an individual binary within the next decade \citep{rosado_expected_properties,Mingarelli:2017, Luke_single_source}.  Both the detection of individual binaries, or their non-detection, provides an additional constraint on SMBHB populations beyond the GWB alone.  In addition to the astrophysical interpretation of the GWB (\citealt{nanograv_15yr_astro}), a followup analysis that explicitly incorporates individual binary upper-limits is underway.

Given the large measurement and modeling uncertainties of the GWB spectrum, detecting an individual binary would also be the most reliable indicator that the background is indeed due to binaries instead of other cosmological sources \citep{nanograv_15yr_new_physics}. These prospects motivate the continuous improvement of our search algorithms such as making them more efficient \citep{QuickCW}, searching for multiple binaries simultaneously (see e.g.,~\citealt{babak_sesana_multiple_cw,BayesHopper}), searching for a correlated background along with the individual binary, searching for eccentric binaries (see e.g.,~\citealt{steve_eccbin, abhimanyu_eccbin_2020, abhimanyu_eccbin_2022}), and many more.

In addition, SMBHBs also produce bright electromagnetic emission \citep{2022LRR....25....3B}, making them exceptional sources for multi-messenger observations \citep{Kelley_white_paper_2019,Charisi+2022}. Multi-messenger analyses can boost the detectability of SMBHBs \citep{tingting2021}, improve the parameter estimation \citep{tingting2023} and improve the GW limits by an order of magnitude \citep{caitlin_3c66b}. A targeted search looking for GW signals in the NANOGrav 15-year dataset from electromagnetic SMBHB candidates is currently underway.

\section{Acknowledgements}
\label{sec:acks}
\textit{Author contributions:}
An alphabetical-order author list was used for this paper in recognition of the fact that a large, decade timescale project such as NANOGrav is necessarily the result of the work of many people. All authors contributed to the activities of the NANOGrav collaboration leading to the work presented here, and reviewed the manuscript, text, and figures prior to the paper’s submission. Additional specific contributions to this paper are as follows.
G.A., A.A., A.M.A., Z.A., P.T.B., P.R.B., H.T.C., K.C., M.E.D., P.B.D., T.D., E.C.F., W.F., E.F., G.E.F., N.G., P.A.G., J.G., D.C.G., J.S.H., R.J.J., M.L.J., D.L.K., M.K., M.T.L., D.R.L., J.L., R.S.L., A.M., M.A.M., N.M., B.W.M., C.N., D.J.N., T.T.P., B.B.P.P., N.S.P., H.A.R., S.M.R., P.S.R., A.S., C.S., B.J.S., I.H.S., K.S., A.S., J.K.S., and H.M.W.~developed the 15-year data set through a combination of observations, arrival time calculations, data checks and refinements, and timing model development and analysis; additional specific contributions to the data set are summarized in NG15 \citet{nanograv_15yr_dataset}.
B.B.~and N.J.C.~coordinated the analysis and paper writing. B.B., N.J.C., and M.C.D.~developed the new analysis method, and B.B.~and M.C.D.~implemented and maintains the \texttt{QuickCW} code used for the analysis. B.B., R.C., M.C.D., K.D.O., P.P., and J.T.~performed analysis for the project, including exploratory runs. B.B., S.C., and C.A.W. updated the pulsar distance priors. S.H.~performed all reweighing analysis to account for correlated noise. A.M.A., B.B., D.L.K., and P.M.~examined the covariance with the binary model of PSR J1713+0747. B.B.~and J.S.H.~explored the use of advanced noise models. M.C.~and P.P.~compared the localization volume of the high-frequency candidate with the NANOGrav galaxy catalog. L.Z.K.~developed the method of calculating number density upper limits. J.S.H., A.B.~and J.S.K.~served as the analysis review team. B.B.~produced the figures. B.B., M.C., N.J.C., L.Z.K.~and K.D.O.~contributed text to the manuscript. P.T.B., M.C., T.D., T.J.W.L., K.D.O., and R.V.H.~provided valuable feedback on the manuscript.

\textit{Acknowledgements.}
The NANOGrav collaboration receives support from National Science Foundation (NSF) Physics Frontiers Center award numbers 1430284 and 2020265, the Gordon and Betty Moore Foundation, NSF AccelNet award number 2114721, an NSERC Discovery Grant, and CIFAR. The Arecibo Observatory is a facility of the NSF operated under cooperative agreement (AST-1744119) by the University of Central Florida (UCF) in alliance with Universidad Ana G. M{\'e}ndez (UAGM) and Yang Enterprises (YEI), Inc. The Green Bank Observatory is a facility of the NSF operated under cooperative agreement by Associated Universities, Inc. The National Radio Astronomy Observatory is a facility of the NSF operated under cooperative agreement by Associated Universities, Inc.
L.B. acknowledges support from the National Science Foundation under award AST-1909933 and from the Research Corporation for Science Advancement under Cottrell Scholar Award No. 27553.
P.R.B. is supported by the Science and Technology Facilities Council, grant number ST/W000946/1.
S.B. gratefully acknowledges the support of a Sloan Fellowship, and the support of NSF under award \#1815664.
The work of R.C., N.La., X.S., and J.T. is partly supported by the George and Hannah Bolinger Memorial Fund in the College of Science at Oregon State University.
M.C., P.P., and S.R.T. acknowledge support from NSF AST-2007993.
M.C. and N.S.P. were supported by the Vanderbilt Initiative in Data Intensive Astrophysics (VIDA) Fellowship.
Support for this work was provided by the NSF through the Grote Reber Fellowship Program administered by Associated Universities, Inc./National Radio Astronomy Observatory.
Support for H.T.C. is provided by NASA through the NASA Hubble Fellowship Program grant \#HST-HF2-51453.001 awarded by the Space Telescope Science Institute, which is operated by the Association of Universities for Research in Astronomy, Inc., for NASA, under contract NAS5-26555.
K.C. is supported by a UBC Four Year Fellowship (6456).
M.E.D. acknowledges support from the Naval Research Laboratory by NASA under contract S-15633Y.
T.D. and M.T.L. are supported by an NSF Astronomy and Astrophysics Grant (AAG) award number 2009468.
E.C.F. is supported by NASA under award number 80GSFC21M0002.
G.E.F., S.C.S., and S.J.V. are supported by NSF award PHY-2011772.
The Flatiron Institute is supported by the Simons Foundation.
S.H. is supported by the National Science Foundation Graduate Research Fellowship under Grant No. DGE-1745301.
A.D.J. and M.V. acknowledge support from the Caltech and Jet Propulsion Laboratory President's and Director's Research and Development Fund.
A.D.J. acknowledges support from the Sloan Foundation.
N.La. acknowledges the support from Larry W. Martin and Joyce B. O'Neill Endowed Fellowship in the College of Science at Oregon State University.
Part of this research was carried out at the Jet Propulsion Laboratory, California Institute of Technology, under a contract with the National Aeronautics and Space Administration (80NM0018D0004).
D.R.L. and M.A.M. are supported by NSF \#1458952.
M.A.M. is supported by NSF \#2009425.
C.M.F.M. was supported in part by the National Science Foundation under Grants No. NSF PHY-1748958 and AST-2106552.
A.Mi. is supported by the Deutsche Forschungsgemeinschaft under Germany's Excellence Strategy - EXC 2121 Quantum Universe - 390833306.
The Dunlap Institute is funded by an endowment established by the David Dunlap family and the University of Toronto.
K.D.O. was supported in part by NSF Grant No. 2207267.
K.D.O. acknowledges the Tufts University High Performance Compute Cluster (https://it.tufts.edu/high-performance-computing) which was utilized for some of the research reported in this paper.
T.T.P. acknowledges support from the Extragalactic Astrophysics Research Group at E\"{o}tv\"{o}s Lor\'{a}nd University, funded by the E\"{o}tv\"{o}s Lor\'{a}nd Research Network (ELKH), which was used during the development of this research.
S.M.R. and I.H.S. are CIFAR Fellows.
Portions of this work performed at NRL were supported by ONR 6.1 basic research funding.
J.D.R. also acknowledges support from start-up funds from Texas Tech University.
J.S. is supported by an NSF Astronomy and Astrophysics Postdoctoral Fellowship under award AST-2202388, and acknowledges previous support by the NSF under award 1847938.
S.R.T. acknowledges support from an NSF CAREER award \#2146016.
C.U. acknowledges support from BGU (Kreitman fellowship), and the Council for Higher Education and Israel Academy of Sciences and Humanities (Excellence fellowship).
C.A.W. acknowledges support from CIERA, the Adler Planetarium, and the Brinson Foundation through a CIERA-Adler postdoctoral fellowship.
O.Y. is supported by the National Science Foundation Graduate Research Fellowship under Grant No. DGE-2139292.
This work was conducted in part using the resources of the Advanced Computing Center for Research and Education (ACCRE) at Vanderbilt University, Nashville, TN.

\facilities{Arecibo, GBT, VLA}

\software{
\texttt{QuickCW} \citep{QuickCW_code},
\texttt{enterprise} \citep{enterprise},
\texttt{enterprise\textunderscore extensions} \citep{ee},
\texttt{libstempo} \citep{libstempo},
\texttt{tempo} \citep{tempo},
\texttt{tempo2} \citep{tempo2},
\texttt{PINT} \citep{pint},
\texttt{matplotlib} \citep{matplotlib},
\texttt{astropy} \citep{astropy, astropy:2013},
\texttt{cosmopy} \citep{cosmopy},
\texttt{healpy} \citep{healpy},  
\texttt{HEALPix} \citep{healpix}
}

\appendix
\section{Software Consistency Checks and Reproducibility}
\label{sec:code_review}

To help make our results easily reproducible, we prepared tutorials that give all technical details needed to verify our main results \citep{nanograv_15yr_cw_sup_mat}. To ensure the credibility of our results, we also carried out several tests on the \texttt{QuickCW} software used to produce the results presented in this paper. We followed procedures outlined in \citet{code_review}, and carried out the following tests, all of which indicated \texttt{QuickCW} was working as intended:
\begin{enumerate}
    \item We compared the likelihood values calculated by \texttt{QuickCW} with the ones directly from the \texttt{enterprise} software package. We obtained good agreement throughout an entire MCMC run, with differences consistent with the expected numerical precision.
    \item We performed prior recovery tests on all sampled parameters. We found that when setting the likelihood to a constant the sampler recovered the priors ensuring sufficient exploration of the entire parameter space.
    \item We carried out a p-p test, which ensures that the sampler gives unbiased estimates of model parameters. We created 40 realizations of a reduced dataset with the 10 longest-timed pulsars with simulated noise made to resemble the real data and a simulated binary with parameters drawn from their priors. Then we analyzed each dataset with a lightweight version of the analysis we used for the real data (significantly fewer iterations). These differences compared to production settings were necessary to make this test computationally feasible. In Figure \ref{fig:pp-plot} we show a so called p-p plot, which shows the cumulative fraction of realizations where the true value of the parameter is at some p-value in the recovered posterior. A diagonal line would mean perfect unbiased recovery, and the gray bands show the 68.3/95.5/99.7\% confidence intervals expected from the finite number of simulations done. We can see that all curves stay within the 99.7\% confidence interval, indicating unbiased parameter recovery.
\end{enumerate}

\begin{figure}[htbp]
    \centering
    \includegraphics[width = 0.5\columnwidth]{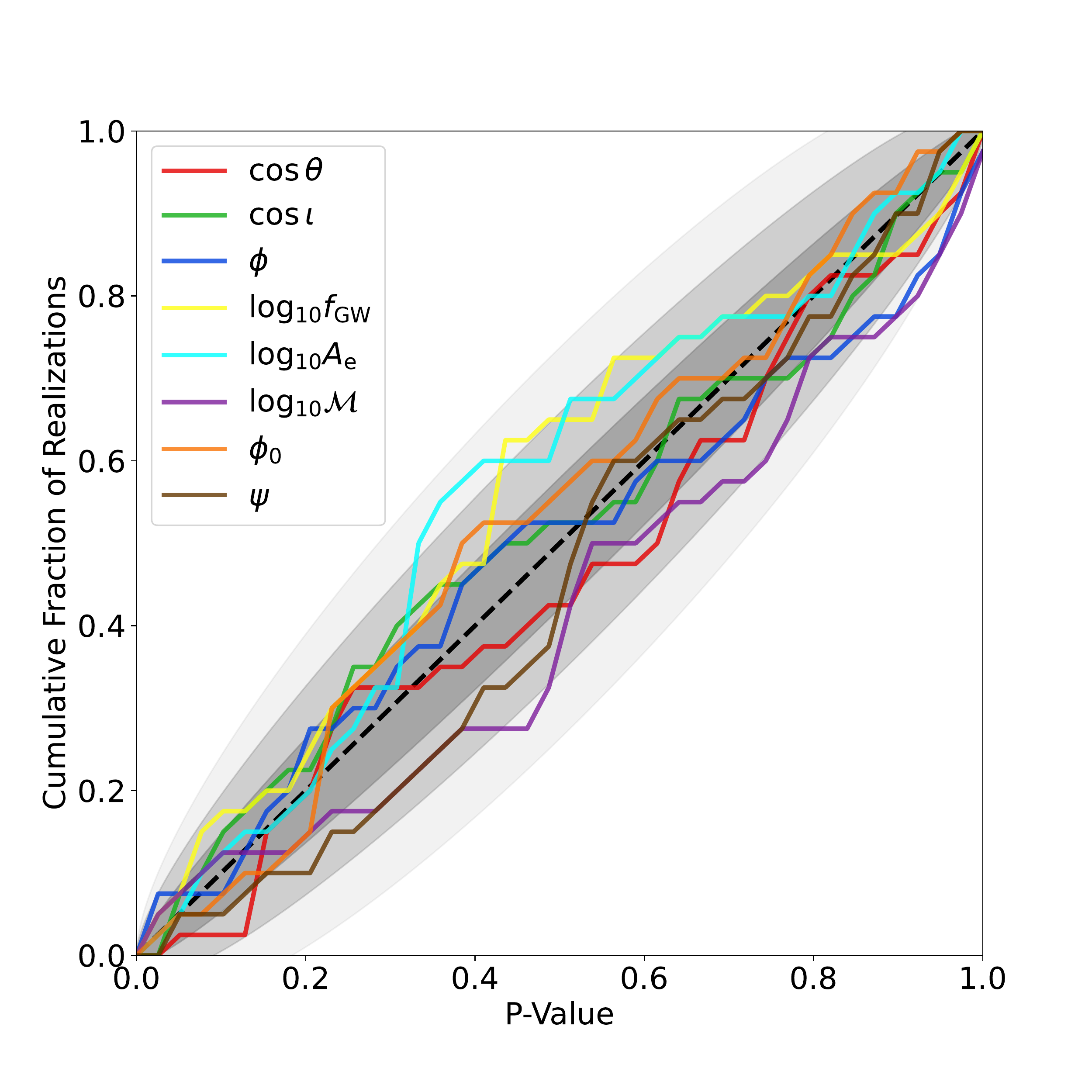}
    \caption{P-P plot showing unbiased recovery of binary parameters. Plotted for eight binary parameters is the cumulative fraction of realizations where the true value of the parameter was at a given percentile of the recovered posterior. The dashed diagonal line indicates perfect recovery, while the three shaded regions show 68.3/95.5/99.7\% confidence intervals based on the number of realizations. All curves stay within these bands demonstrating unbiased parameter recovery.}
    \label{fig:pp-plot}
\end{figure}

\hspace{1cm}

\section{Pulsar Distance Values}
\label{sec:psr_dist}

\begin{longtable*}[c]{c|cccp{0.55\linewidth}}
    \caption{Compiled pulsar distance values and uncertainties for each pulsar used for the analysis presented in this paper. The second column indicates whether the distance value is from dispersion measure (DM) or parallax (PX) measurements.}
    \label{tab:psr_dists}
    \\
    \hline
    Pulsar & Prior & Distance [kpc] & Error [kpc] & References\\
    \hline
    \hline
    \endhead
    \hline
    \endfoot
B1855+09 & PX & 1.18 & 0.12 & \citet{http://adsabs.harvard.edu/abs/2009MNRAS.400..951V, http://adsabs.harvard.edu/abs/1994ApJ...428..713K, nanograv_15yr_dataset}\\ 
B1937+21 & PX & 3.10 & 0.20 & \citet{https://doi.org/10.1093/mnras/stac3725, nanograv_15yr_dataset}\\ 
B1953+29 & DM & 4.64 & 0.93 & \\ 
J0023+0923 & PX & 1.02 & 0.11 & \citet{nanograv_15yr_dataset}\\ 
J0030+0451 & PX & 0.3296 & 0.0036 & \citet{https://doi.org/10.1093/mnras/stac3725, http://adsabs.harvard.edu/abs/2006ApJ...642.1012L, nanograv_15yr_dataset}\\ 
J0340+4130 & DM & 1.71 & 0.34 & \\ 
J0406+3039 & DM & 1.72 & 0.34 & \\ 
J0437$-$4715 & PX & 0.1549 & 0.0010 & \citet{ttps://ui.adsabs.harvard.edu/abs/2018ApJ...864...26J/, http://adsabs.harvard.edu/abs/2008ApJ...685L..67D, http://adsabs.harvard.edu/abs/2008ApJ...679..675V, http://adsabs.harvard.edu/abs/2006MNRAS.369.1502H, http://adsabs.harvard.edu/abs/2001Natur.412..158V, http://adsabs.harvard.edu/abs/1997ApJ...478L..95S, nanograv_15yr_dataset}\\ 
J0509+0856 & DM & 1.45 & 0.29 & \\ 
J0557+1551 & DM & 2.87 & 0.57 & \\ 
J0605+3757 & DM & 0.70 & 0.14 & \\ 
J0610$-$2100 & PX & 1.38 & 0.14 & \citet{https://doi.org/10.1093/mnras/stac3725}\\ 
J0613$-$0200 & PX & 1.07 & 0.10 & \citet{http://adsabs.harvard.edu/abs/2006MNRAS.369.1502H, nanograv_15yr_dataset}\\ 
J0636+5128 & PX & 1.27 & 0.29 & \citet{nanograv_15yr_dataset}\\ 
J0645+5158 & PX & 1.37 & 0.19 & \citet{http://adsabs.harvard.edu/abs/2014ApJ...791...67S, nanograv_15yr_dataset}\\ 
J0709+0458 & DM & 1.80 & 0.36 & \\ 
J0740+6620 & PX & 0.94 & 0.19 & \citet{nanograv_15yr_dataset}\\ 
J0931$-$1902 & DM & 1.88 & 0.38 & \\ 
J1012+5307 & PX & 0.862 & 0.022 & \citet{https://doi.org/10.1093/mnras/stac3725, ttps://ui.adsabs.harvard.edu/abs/2020ApJ...896...85D, http://adsabs.harvard.edu/abs/2009MNRAS.400..805L}\\ 
J1012$-$4235 & DM & 2.50 & 0.50 & \\ 
J1022+1001 & PX & 0.706 & 0.019 & \citet{https://ui.adsabs.harvard.edu/abs/2019ApJ...875..100D, http://adsabs.harvard.edu/abs/2009MNRAS.400..951V, http://adsabs.harvard.edu/abs/2006MNRAS.369.1502H, http://adsabs.harvard.edu/abs/2004MNRAS.355..941H, nanograv_15yr_dataset}\\ 
J1024$-$0719 & PX & 1.080 & 0.042 & \citet{https://doi.org/10.1093/mnras/stac3725, http://dx.doi.org/10.1051/0004-6361/201527847, nanograv_15yr_dataset}\\ 
J1125+7819 & DM & 0.63 & 0.13 & \\ 
J1312+0051 & DM & 0.84 & 0.17 & \\ 
J1453+1902 & DM & 1.15 & 0.23 & \\ 
J1455$-$3330 & PX & 1.01 & 0.22 & \citet{http://dx.doi.org/10.1051/0004-6361/201527847}\\ 
J1600$-$3053 & PX & 1.84 & 0.26 & \citet{nanograv_15yr_dataset}\\ 
J1614$-$2230 & PX & 0.699 & 0.026 & \citet{http://dx.doi.org/10.1051/0004-6361/201527847, http://adsabs.harvard.edu/abs/2013ApJS..208...17A, nanograv_15yr_dataset}\\ 
J1630+3734 & PX & 0.089 & 0.024 & \citet{nanograv_15yr_dataset}\\ 
J1640+2224 & PX & 1.404 & 0.095 & \citet{https://doi.org/10.1093/mnras/stac3725}\\ 
J1643$-$1224 & PX & 0.835 & 0.059 & \citet{https://doi.org/10.1093/mnras/stac3725, http://adsabs.harvard.edu/abs/2009MNRAS.400..951V}\\ 
J1705$-$1903 & DM & 1.62 & 0.32 & \\ 
J1713+0747 & PX & 1.138 & 0.019 & \citet{http://adsabs.harvard.edu/abs/2009MNRAS.400..951V, http://adsabs.harvard.edu/abs/2009ApJ...698..250C, http://adsabs.harvard.edu/abs/2006MNRAS.369.1502H, http://adsabs.harvard.edu/abs/2005ApJ...620..405S, nanograv_15yr_dataset}\\ 
J1719$-$1438 & DM & 1.21 & 0.24 & \\ 
J1730$-$2304 & PX & 0.529 & 0.022 & \citet{https://doi.org/10.1093/mnras/stac3725, http://dx.doi.org/10.1051/0004-6361/201527847, nanograv_15yr_dataset}\\ 
J1738+0333 & PX & 1.64 & 0.10 & \citet{https://doi.org/10.1093/mnras/stac3725, http://adsabs.harvard.edu/abs/2012MNRAS.423.3328F, nanograv_15yr_dataset}\\ 
J1741+1351 & PX & 2.36 & 0.56 & \citet{nanograv_15yr_dataset}\\ 
J1744$-$1134 & PX & 0.4141 & 0.0093 & \citet{http://adsabs.harvard.edu/abs/2009MNRAS.400..951V, http://adsabs.harvard.edu/abs/2006MNRAS.369.1502H, http://adsabs.harvard.edu/abs/1999ApJ...523L.171T, nanograv_15yr_dataset}\\ 
J1745+1017 & DM & 1.27 & 0.25 & \\ 
J1747$-$4036 & DM & 3.50 & 0.70 & \\ 
J1751$-$2857 & DM & 1.11 & 0.22 & \\ 
J1802$-$2124 & DM & 2.96 & 0.59 & \\ 
J1811$-$2405 & DM & 1.79 & 0.36 & \\ 
J1832$-$0836 & PX & 2.00 & 0.47 & \citet{nanograv_15yr_dataset}\\ 
J1843$-$1113 & DM & 1.72 & 0.34 & \\ 
J1853+1303 & PX & 1.91 & 0.17 & \citet{https://doi.org/10.1093/mnras/stac3725, nanograv_15yr_dataset}\\ 
J1903+0327 & DM & 6.49 & 1.30 & \\ 
J1909$-$3744 & PX & 1.159 & 0.013 & \citet{http://adsabs.harvard.edu/abs/2009MNRAS.400..951V, http://adsabs.harvard.edu/abs/2006MNRAS.369.1502H, http://adsabs.harvard.edu/abs/2005ApJ...629L.113J, nanograv_15yr_dataset}\\ 
J1910+1256 & PX & 3.52 & 0.41 & \citet{https://doi.org/10.1093/mnras/stac3725}\\ 
J1911+1347 & DM & 2.08 & 0.42 & \\ 
J1918$-$0642 & PX & 1.44 & 0.11 & \citet{https://doi.org/10.1093/mnras/stac3725, nanograv_15yr_dataset}\\ 
J1923+2515 & PX & 0.94 & 0.21 & \citet{nanograv_15yr_dataset}\\ 
J1944+0907 & PX & 1.38 & 0.36 & \citet{nanograv_15yr_dataset}\\ 
J1946+3417 & DM & 5.12 & 1.02 & \\ 
J2010$-$1323 & PX & 1.94 & 0.41 & \citet{nanograv_15yr_dataset}\\ 
J2017+0603 & DM & 1.57 & 0.31 & \\ 
J2033+1734 & DM & 1.99 & 0.40 & \\ 
J2043+1711 & PX & 1.58 & 0.11 & \citet{nanograv_15yr_dataset}\\ 
J2124$-$3358 & PX & 0.413 & 0.055 & \citet{http://adsabs.harvard.edu/abs/2009MNRAS.400..951V, nanograv_15yr_dataset}\\ 
J2145$-$0750 & PX & 0.624 & 0.022 & \citet{https://ui.adsabs.harvard.edu/abs/2019ApJ...875..100D/, http://adsabs.harvard.edu/abs/2009MNRAS.400..951V, nanograv_15yr_dataset}\\ 
J2214+3000 & DM & 1.54 & 0.31 & \\ 
J2229+2643 & DM & 1.43 & 0.29 & \\ 
J2234+0611 & PX & 1.23 & 0.17 & \citet{nanograv_15yr_dataset}\\ 
J2234+0944 & DM & 1.00 & 0.20 & \\ 
J2302+4442 & DM & 1.18 & 0.24 & \\ 
J2317+1439 & PX & 1.57 & 0.29 & \citet{nanograv_15yr_dataset}\\ 
J2322+2057 & PX & 1.00 & 0.21 & \citet{nanograv_15yr_dataset}\\  
\hline
\end{longtable*}

\section{Effects of Noise Modeling on High-frequency Candidate}
\label{sec:adv_noise}

It is well known that PTA datasets can be prone to having unmodeled high-frequency noise. E.g.,~this is why the number of frequency components used to model the GWB is usually limited to avoid contamination from higher frequencies (see e.g.,~\citealt{nanograv_15yr_gwb, nanograv_15yr_detchar}). It has also been shown that including advanced noise models can help alleviate these problems \citep{ppta_dr2_noise, ipta_dr2_cw, adv_noise}. To test if these models can also help explaining the high-frequency candidate we found, we used advanced noise models for some of the pulsars using the maximum likelihood parameter values of these models derived based on the NANOGrav 12.5-year dataset \citep{adv_noise}. To decide which pulsars to use these models for, we created an ordered list of them based on how informative the given pulsar's pulsar phase posteriors were. This is an indicator of how much they support the presence of the candidate, and can be measured by the Kullback–Leibler divergence between the prior and the posterior. The top nine pulsars in this measure were: J1713+0747, B1937+21, J1909$-$3744, J1012+5307, J2317+1439, J1918$-$0642, J1640+2224, J1741+1351, J0613$-$0200. However, we did not have advanced noise models available for PSR J1918$-$0642, leaving us with eight pulsars to include advanced noise models for. We did three different runs: (1) where only PSR J1713+0747 had an advanced noise model included; (2) where the four top pulsars listed above had advanced noise models included; and (3) where all eight top pulsars listed above had advanced noise models included. The corresponding Bayes factors are shown in Figure \ref{fig:bf_highf} in blue, aqua and purple, respectively. Including advanced noise models decreases the Bayes factors of this high-frequency candidate, but it is still the largest Bayes factor we see in the whole search. The significant variation in the Bayes factors with changes in the noise model for these high frequency candidates suggests that more work needs to be done to understand the noise in this frequency band.

\begin{figure}[htbp]
    \centering
    \includegraphics[width = 0.8\columnwidth]{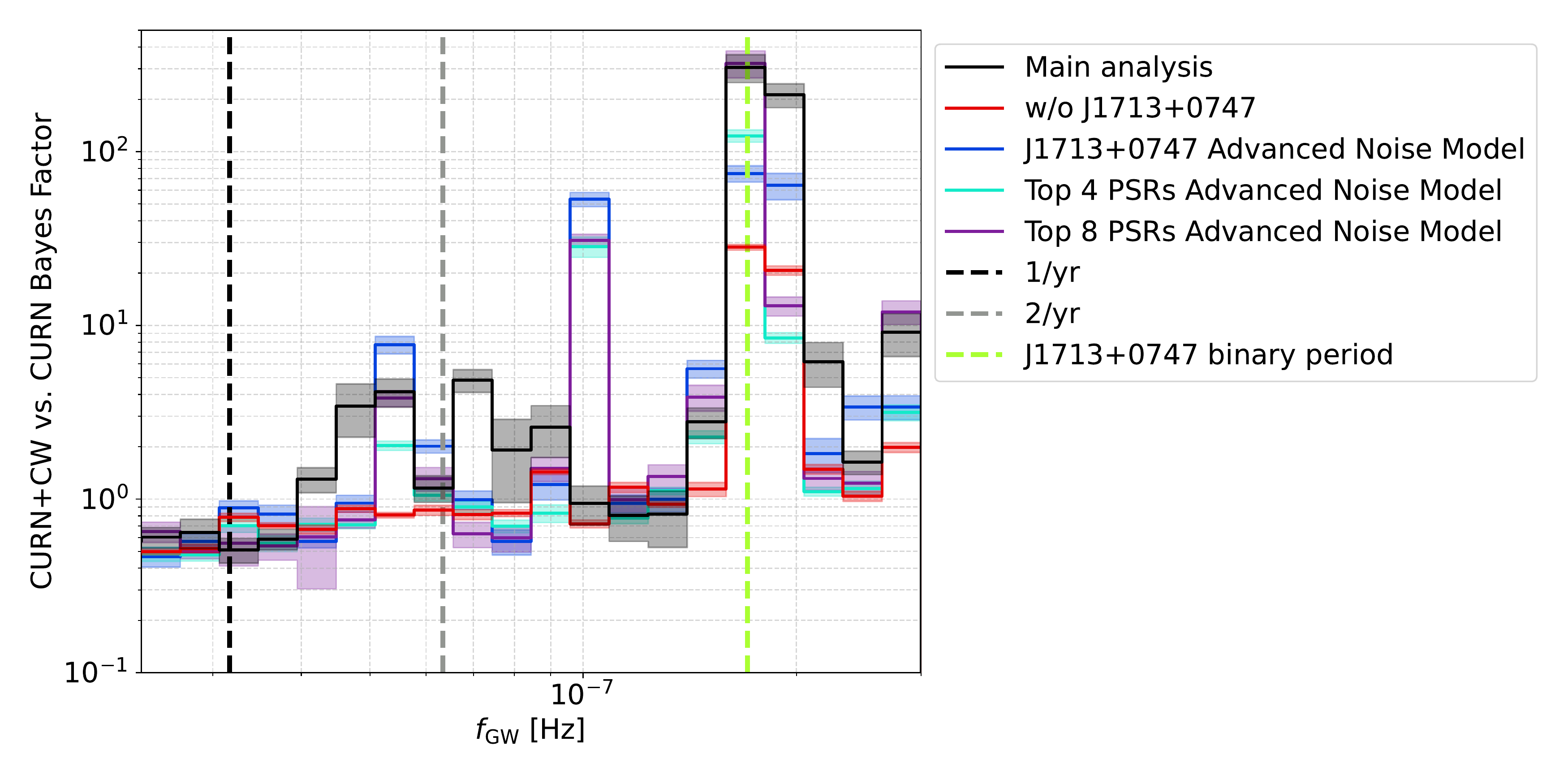}
    \caption{Savage-Dickey Bayes factors for the CW+CURN model versus the CURN model as a function of frequency (black). Also shown are Bayes factors when excluding PSR J1713+0747 (red); when advanced noise models are used for PSR J1713+0747 (blue); for PSRs J1713+0747, B1937+21, J1909$-$3744, and J1012+5307 (aqua); and for PSRs J1713+0747, B1937+21, J1909$-$3744, J1012+5307, J2317+1439, J1640+2224, J1741+1351, and J0613$-$0200 (purple). Shaded regions show the 1-$\sigma$ uncertainties.}
    \label{fig:bf_highf}
\end{figure}

\section{Effects of Prior Choice and Correlations on Upper Limits}
\label{sec:UL_prior}

In Figure \ref{fig:ul} we show the amplitude upper limits as a function of frequency we get with uniform (red) and log-uniform (orange) amplitude priors. As expected, the upper limit using the uniform prior is consistently higher than the one using the log-uniform prior. Note that around the low-frequency candidate discussed in Section \ref{ssec:lowf_candidate} ($\sim$ 4 nHz) and around the high-frequency candidate discussed in \ref{ssec:highf_candidate} ($\sim$ 200 nHz), the two curves are closer to each other. This indicates that once the presence of a signal is preferred and the posterior is likelihood-dominated, the amplitude prior choice has a smaller effect on the upper limits. As we can see, the upper limits depend on our prior choices. However, the uniform amplitude prior has the desirable property that the upper limit is independent of the lower and upper prior bounds used, since the posterior diminishes towards both limits. This is not true for the log-uniform prior, for which the posterior often extends all the way to the lower prior bound, resulting in an upper limit that depends on the chosen lower prior bound.

Figure \ref{fig:ul} also shows the upper limits after taking the HD-correlations into account with resampling. Note that the correlations make a difference for the log-uniform amplitude prior run, but not for the uniform amplitude prior run. This is most likely due to the fact that the latter biases the CW amplitude high, so that it is less affected by the GWB, and whether that has correlations or not makes less difference. We also show the amplitude and frequency of the low-frequency candidate, which is consistent with the upper limits regardless of the priors used.

\begin{figure}[h]
    \centering
    \includegraphics[width = 0.6\columnwidth]{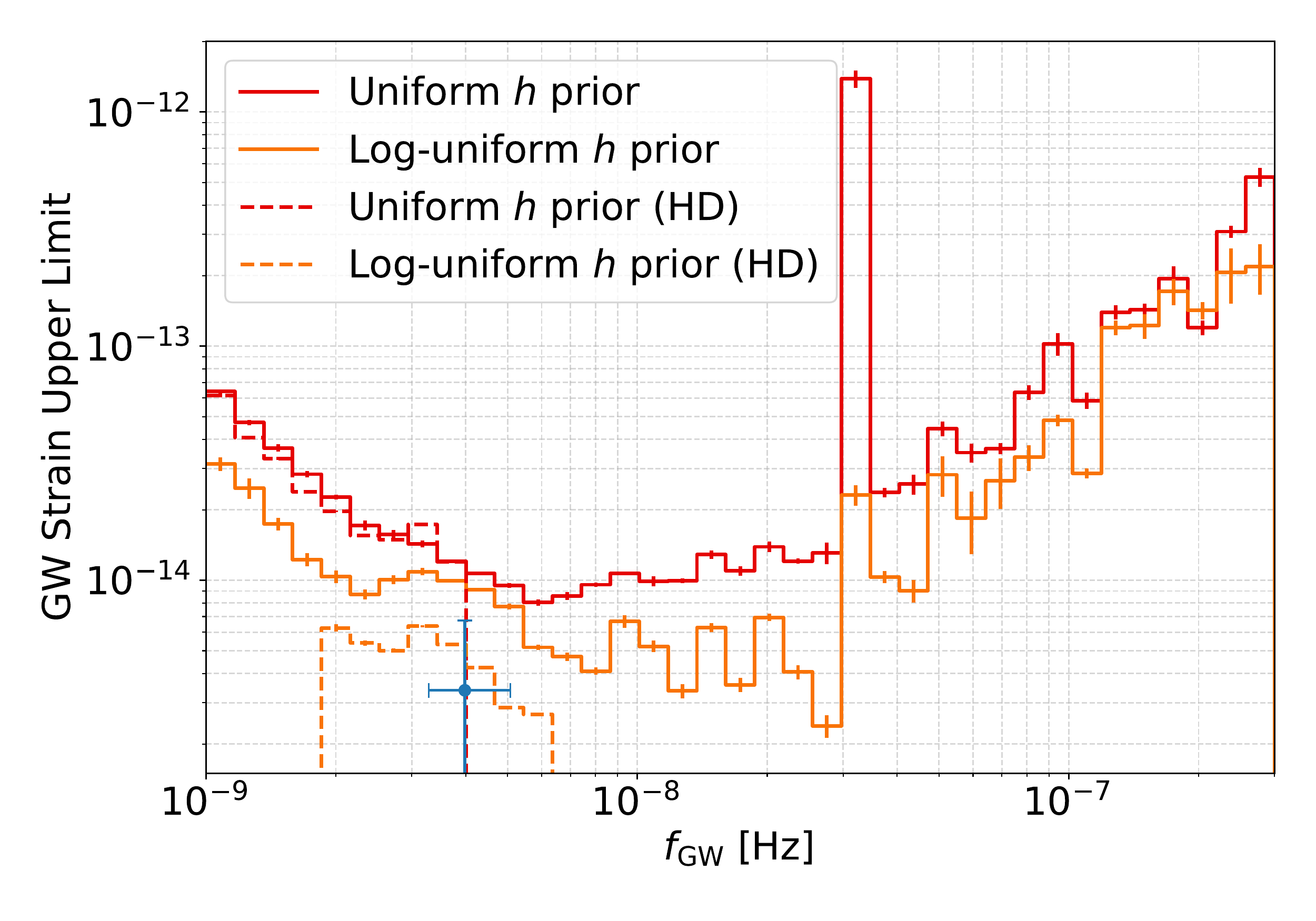}
    \caption{All-sky CW strain 95\% upper limits using a uniform (red) and log-uniform (orange) prior on the amplitude. Dashed lines show the upper limits we get if we take HD correlations into account with reweighing. These are only calculated over the range of frequencies where we had enough samples for the reweighing. The blue marker indicates the frequency and the amplitude of the low-frequency candidate discussed in Section \ref{ssec:lowf_candidate} with 68\% credible intervals.}
    \label{fig:ul}
\end{figure}

\section{Comparison with 12.5-year results}
\label{sec:UL_12p5yr}

As we have seen in Figure \ref{fig:ul_progress}, the GW strain upper limits based on the 15-year dataset are improved at the lowest frequencies compared to the 12.5-year dataset, but not at higher frequencies. In particular, the 12.5-year dataset is more sensitive than the 15-year dataset around 8 nHz. However, the comparison is complicated due to the differences in analysis techniques used. To better understand the source of these differences, we reanalyzed the 12.5-year dataset with the techniques used in this paper. The resulting upper limit curve is shown in Figure \ref{fig:ul_12p5yr} (purple), along with the 15-year result (red) and the original 12.5-year result (blue). We can see that while there are small differences, the QuickCW analysis of the 12.5-year dataset is in agreement with the original 12.5-year result.

\begin{figure}[h]
    \centering
    \includegraphics[width = 0.6\columnwidth]{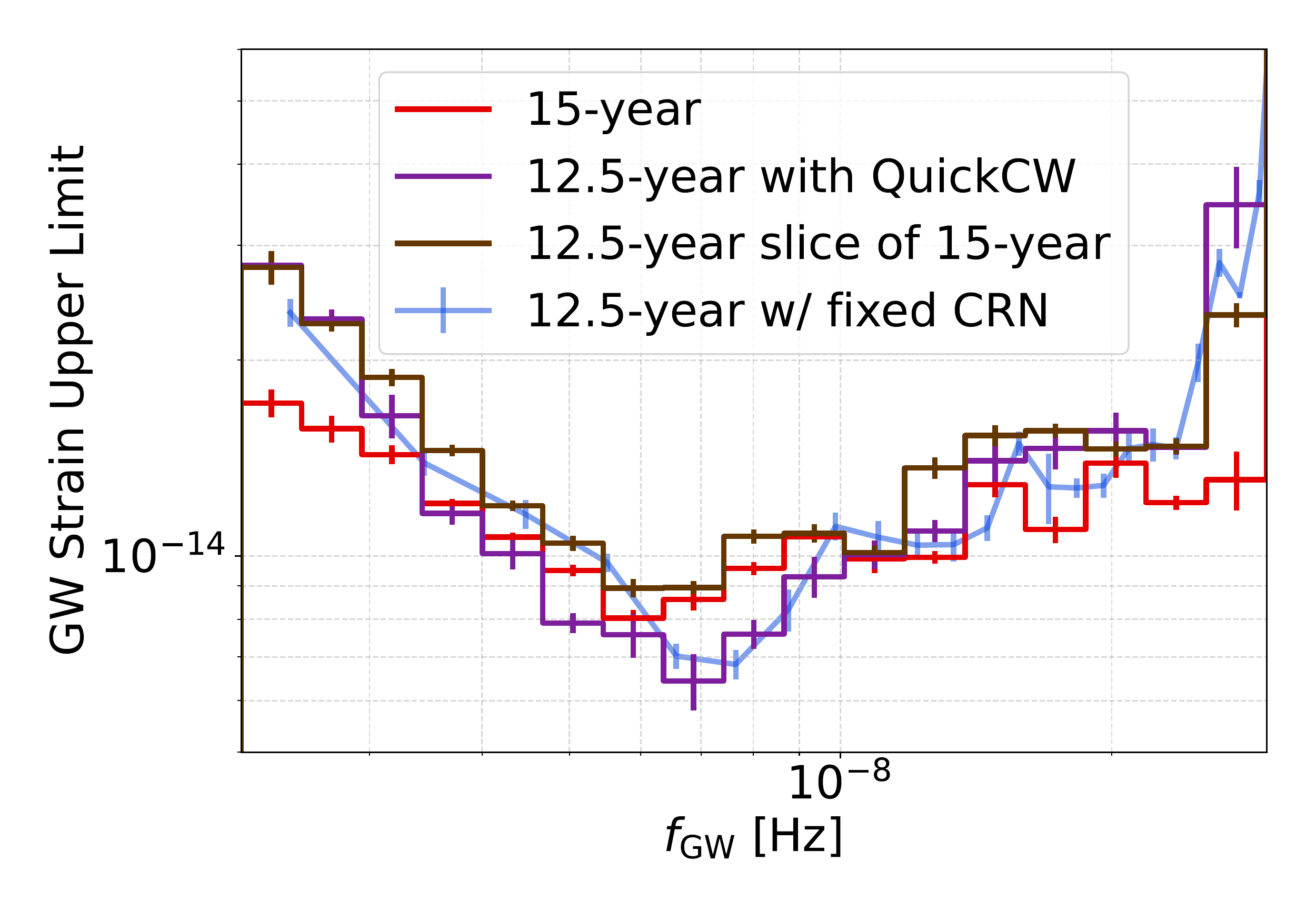}
    \caption{All-sky CW strain 95\% upper limits from the NANOGrav 12.5-year dataset with QuickCW (purple). We also show the official 12.5-year upper limits (blue), which despite the different methodology is consistent with the QuickCW results. Also shown for comparison is the 15-year upper limit curve (red).  We also analyzed the 12.5-year slice of the 15-year dataset (brown), which gives results similar to the 12.5-year dataset at low and high frequencies, but follows the 15-year results more closely at moderate frequencies.}
    \label{fig:ul_12p5yr}
\end{figure}

We also analyzed the 12.5-year time-slice of the 15-year dataset. Time-slicing is a useful technique for understanding how results depend on the data included in the analysis \citep{11yr_slice}, while keeping the timing solutions and white noise levels determined based on the full dataset. Our time slice is based on the 15-year dataset, but we only include pulsars that were already present in the 12.5-year dataset, and we cut out any of their data that was recorded after the end date of the 12.5-year dataset. We fixed these pulsars' white noise parameters based on the noise analysis on the full 15-year data. The results are shown on Figure \ref{fig:ul_12p5yr} (brown). We can see that this curve follows the 12.5-year curves very closely at low ($\lesssim 3$ nHz) and high ($\gtrsim 10$ nHz) frequencies. However, at frequencies around 8 nHz, this analysis gives results more similar to the full 15-year analysis. This indicates that the difference in the upper limit at these frequencies between the 12.5-year and 15-year datasets is not simply due to the additional data points. Instead, it hints to these differences being due to some difference in the processing of the dataset, highlighting that new datasets are not just incremental updates of previous datasets, but instead complete reanalyses of all available data.

\bibliographystyle{aasjournal}
\bibliography{cw15yr}

\begin{thebibliography}{}
\expandafter\ifx\csname natexlab\endcsname\relax\def\natexlab#1{#1}\fi
\providecommand{\url}[1]{\href{#1}{#1}}
\providecommand{\dodoi}[1]{doi:~\href{http://doi.org/#1}{\nolinkurl{#1}}}
\providecommand{\doeprint}[1]{\href{http://ascl.net/#1}{\nolinkurl{http://ascl.net/#1}}}
\providecommand{\doarXiv}[1]{\href{https://arxiv.org/abs/#1}{\nolinkurl{https://arxiv.org/abs/#1}}}

\bibitem[{{Abdo} {et~al.}(2013){Abdo}, {Ajello}, {Allafort}, {Baldini},
  {Ballet}, {Barbiellini}, {Baring}, {Bastieri}, {Belfiore}, {Bellazzini},
  {Bhattacharyya}, {Bissaldi}, {Bloom}, {Bonamente}, {Bottacini}, {Brandt},
  {Bregeon}, {Brigida}, {Bruel}, {Buehler}, {Burgay}, {Burnett}, {Busetto},
  {Buson}, {Caliandro}, {Cameron}, {Camilo}, {Caraveo}, {Casandjian}, {Cecchi},
  {{\c{C}}elik}, {Charles}, {Chaty}, {Chaves}, {Chekhtman}, {Chen}, {Chiang},
  {Chiaro}, {Ciprini}, {Claus}, {Cognard}, {Cohen-Tanugi}, {Cominsky},
  {Conrad}, {Cutini}, {D'Ammando}, {de Angelis}, {DeCesar}, {De Luca}, {den
  Hartog}, {de Palma}, {Dermer}, {Desvignes}, {Digel}, {Di Venere}, {Drell},
  {Drlica-Wagner}, {Dubois}, {Dumora}, {Espinoza}, {Falletti}, {Favuzzi},
  {Ferrara}, {Focke}, {Franckowiak}, {Freire}, {Funk}, {Fusco}, {Gargano},
  {Gasparrini}, {Germani}, {Giglietto}, {Giommi}, {Giordano}, {Giroletti},
  {Glanzman}, {Godfrey}, {Gotthelf}, {Grenier}, {Grondin}, {Grove},
  {Guillemot}, {Guiriec}, {Hadasch}, {Hanabata}, {Harding}, {Hayashida},
  {Hays}, {Hessels}, {Hewitt}, {Hill}, {Horan}, {Hou}, {Hughes}, {Jackson},
  {Janssen}, {Jogler}, {J{\'o}hannesson}, {Johnson}, {Johnson}, {Johnson},
  {Johnson}, {Johnston}, {Kamae}, {Kataoka}, {Keith}, {Kerr}, {Kn{\"o}dlseder},
  {Kramer}, {Kuss}, {Lande}, {Larsson}, {Latronico}, {Lemoine-Goumard},
  {Longo}, {Loparco}, {Lovellette}, {Lubrano}, {Lyne}, {Manchester}, {Marelli},
  {Massaro}, {Mayer}, {Mazziotta}, {McEnery}, {McLaughlin}, {Mehault},
  {Michelson}, {Mignani}, {Mitthumsiri}, {Mizuno}, {Moiseev}, {Monzani},
  {Morselli}, {Moskalenko}, {Murgia}, {Nakamori}, {Nemmen}, {Nuss}, {Ohno},
  {Ohsugi}, {Orienti}, {Orlando}, {Ormes}, {Paneque}, {Panetta}, {Parent},
  {Perkins}, {Pesce-Rollins}, {Pierbattista}, {Piron}, {Pivato}, {Pletsch},
  {Porter}, {Possenti}, {Rain{\`o}}, {Rando}, {Ransom}, {Ray}, {Razzano},
  {Rea}, {Reimer}, {Reimer}, {Renault}, {Reposeur}, {Ritz}, {Romani}, {Roth},
  {Rousseau}, {Roy}, {Ruan}, {Sartori}, {Saz Parkinson}, {Scargle}, {Schulz},
  {Sgr{\`o}}, {Shannon}, {Siskind}, {Smith}, {Spandre}, {Spinelli}, {Stappers},
  {Strong}, {Suson}, {Takahashi}, {Thayer}, {Thayer}, {Theureau}, {Thompson},
  {Thorsett}, {Tibaldo}, {Tibolla}, {Tinivella}, {Torres}, {Tosti}, {Troja},
  {Uchiyama}, {Usher}, {Vandenbroucke}, {Vasileiou}, {Venter}, {Vianello},
  {Vitale}, {Wang}, {Weltevrede}, {Winer}, {Wolff}, {Wood}, {Wood}, {Wood}, \&
  {Yang}}]{http://adsabs.harvard.edu/abs/2013ApJS..208...17A}
{Abdo}, A.~A., {Ajello}, M., {Allafort}, A., {et~al.} 2013, \apjs, 208, 17,
  \dodoi{10.1088/0067-0049/208/2/17}

\bibitem[{{Afzal} {et~al.}(2023)}]{nanograv_15yr_new_physics}
{Afzal}, A., {et~al.} 2023, \apjl

\bibitem[{{Agazie} {et~al.}(2023{\natexlab{a}})}]{nanograv_15yr_dataset}
{Agazie}, G., {et~al.} 2023{\natexlab{a}}, \apjl

\bibitem[{{Agazie} {et~al.}(2023{\natexlab{b}})}]{nanograv_15yr_gwb}
---. 2023{\natexlab{b}}, \apjl

\bibitem[{{Agazie} {et~al.}(2023{\natexlab{c}})}]{nanograv_15yr_astro}
---. 2023{\natexlab{c}}, \apjl

\bibitem[{{Agazie} {et~al.}(2023{\natexlab{d}})}]{nanograv_15yr_detchar}
---. 2023{\natexlab{d}}, \apjl

\bibitem[{{Aggarwal} {et~al.}(2019){Aggarwal}, {Arzoumanian}, {Baker},
  {Brazier}, {Brinson}, {Brook}, {Burke-Spolaor}, {Chatterjee}, {Cordes},
  {Cornish}, {Crawford}, {Crowter}, {Cromartie}, {DeCesar}, {Demorest},
  {Dolch}, {Ellis}, {Ferdman}, {Ferrara}, {Fonseca}, {Garver-Daniels},
  {Gentile}, {Hazboun}, {Holgado}, {Huerta}, {Islo}, {Jennings}, {Jones},
  {Jones}, {Kaiser}, {Kaplan}, {Kelley}, {Key}, {Lam}, {Lazio}, {Levin},
  {Lorimer}, {Luo}, {Lynch}, {Madison}, {McLaughlin}, {McWilliams},
  {Mingarelli}, {Ng}, {Nice}, {Pennucci}, {Pol}, {Ransom}, {Ray}, {Siemens},
  {Simon}, {Spiewak}, {Stairs}, {Stinebring}, {Stovall}, {Swiggum}, {Taylor},
  {Turner}, {Vallisneri}, {van Haasteren}, {Vigeland }, {Witt}, {Zhu}, \& {(The
  NANOGrav Collaboration}}]{nanograv_11yr_cw}
{Aggarwal}, K., {Arzoumanian}, Z., {Baker}, P.~T., {et~al.} 2019, \apj, 880,
  116, \dodoi{10.3847/1538-4357/ab2236}

\bibitem[{{Alam} {et~al.}(2021){Alam}, {Arzoumanian}, {Baker}, {Blumer},
  {Bohler}, {Brazier}, {Brook}, {Burke-Spolaor}, {Caballero}, {Camuccio},
  {Chamberlain}, {Chatterjee}, {Cordes}, {Cornish}, {Crawford}, {Cromartie},
  {Decesar}, {Demorest}, {Dolch}, {Ellis}, {Ferdman}, {Ferrara}, {Fiore},
  {Fonseca}, {Garcia}, {Garver-Daniels}, {Gentile}, {Good}, {Gusdorff},
  {Halmrast}, {Hazboun}, {Islo}, {Jennings}, {Jessup}, {Jones}, {Kaiser},
  {Kaplan}, {Kelley}, {Key}, {Lam}, {Lazio}, {Lorimer}, {Luo}, {Lynch},
  {Madison}, {Maraccini}, {McLaughlin}, {Mingarelli}, {Ng}, {Nguyen}, {Nice},
  {Pennucci}, {Pol}, {Ramette}, {Ransom}, {Ray}, {Shapiro-Albert}, {Siemens},
  {Simon}, {Spiewak}, {Stairs}, {Stinebring}, {Stovall}, {Swiggum}, {Taylor},
  {Tripepi}, {Vallisneri}, {Vigeland}, {Witt}, {Zhu}, \& {Nanograv
  Collaboration}}]{nanograv_12p5yr_data}
{Alam}, M.~F., {Arzoumanian}, Z., {Baker}, P.~T., {et~al.} 2021, \apjs, 252, 4,
  \dodoi{10.3847/1538-4365/abc6a0}

\bibitem[{Allen(2023)}]{Allen:2022dzg}
Allen, B. 2023, Phys. Rev. D, 107, 043018, \dodoi{10.1103/PhysRevD.107.043018}

\bibitem[{{Antoniadis} {et~al.}(2023{\natexlab{a}})}]{epta_dr2_dataset}
{Antoniadis}, J., {et~al.} 2023{\natexlab{a}}, \aap

\bibitem[{{Antoniadis} {et~al.}(2023{\natexlab{b}})}]{epta_dr2_gwb}
---. 2023{\natexlab{b}}, \aap

\bibitem[{{Arzoumanian} {et~al.}(2014){Arzoumanian}, {Brazier},
  {Burke-Spolaor}, {Chamberlin}, {Chatterjee}, {Cordes}, {Demorest}, {Deng},
  {Dolch}, {Ellis}, {Ferdman}, {Garver-Daniels}, {Jenet}, {Jones}, {Kaspi},
  {Koop}, {Lam}, {Lazio}, {Lommen}, {Lorimer}, {Luo}, {Lynch}, {Madison},
  {McLaughlin}, {McWilliams}, {Nice}, {Palliyaguru}, {Pennucci}, {Ransom},
  {Sesana}, {Siemens}, {Stairs}, {Stinebring}, {Stovall}, {Swiggum},
  {Vallisneri}, {van Haasteren}, {Wang}, {Zhu}, \& {NANOGrav
  Collaboration}}]{nanograv_5yr_cw}
{Arzoumanian}, Z., {Brazier}, A., {Burke-Spolaor}, S., {et~al.} 2014, \apj,
  794, 141, \dodoi{10.1088/0004-637X/794/2/141}

\bibitem[{{Arzoumanian} {et~al.}(2020){Arzoumanian}, {Baker}, {Brazier},
  {Brook}, {Burke-Spolaor}, {B{\'e}csy}, {Charisi}, {Chatterjee}, {Cordes},
  {Cornish}, {Crawford}, {Cromartie}, {Crowter}, {Decesar}, {Demorest},
  {Dolch}, {Elliott}, {Ellis}, {Ferdman}, {Ferrara}, {Fonseca},
  {Garver-Daniels}, {Gentile}, {Good}, {Hazboun}, {Islo}, {Jennings}, {Jones},
  {Kaiser}, {Kaplan}, {Kelley}, {Key}, {Lam}, {Lazio}, {Levin}, {Luo}, {Lynch},
  {Madison}, {McLaughlin}, {Mingarelli}, {Ng}, {Nice}, {Pennucci}, {Pol},
  {Ransom}, {Ray}, {Shapiro-Albert}, {Siemens}, {Simon}, {Spiewak}, {Stairs},
  {Stinebring}, {Stovall}, {Swiggum}, {Taylor}, {Vallisneri}, {Vigeland},
  {Witt}, {Zhu}, \& {NANOGrav Collaboration}}]{caitlin_3c66b}
{Arzoumanian}, Z., {Baker}, P.~T., {Brazier}, A., {et~al.} 2020, \apj, 900,
  102, \dodoi{10.3847/1538-4357/ababa1}

\bibitem[{{Arzoumanian} {et~al.}(2021){Arzoumanian}, {Baker}, {Brazier},
  {Brook}, {Burke-Spolaor}, {Becsy}, {Charisi}, {Chatterjee}, {Cordes},
  {Cornish}, {Crawford}, {Cromartie}, {Decesar}, {Demorest}, {Dolch},
  {Elliott}, {Ellis}, {Ferrara}, {Fonseca}, {Garver-Daniels}, {Gentile},
  {Good}, {Hazboun}, {Islo}, {Jennings}, {Jones}, {Kaiser}, {Kaplan}, {Kelley},
  {Key}, {Lam}, {Lazio}, {Luo}, {Lynch}, {Ma}, {Madison}, {McLaughlin},
  {Mingarelli}, {Ng}, {Nice}, {Pennucci}, {Pol}, {Ransom}, {Ray},
  {Shapiro-Albert}, {Siemens}, {Simon}, {Spiewak}, {Stairs}, {Stinebring},
  {Stovall}, {Swiggum}, {Taylor}, {Vallisneri}, {Vigeland}, {Witt}, \&
  {Nanograv Collaboration}}]{Arzoumanian+2021}
---. 2021, \apj, 914, 121, \dodoi{10.3847/1538-4357/abfcd3}

\bibitem[{{Arzoumanian} {et~al.}(2023){Arzoumanian}, {Baker}, {Blecha},
  {Blumer}, {Brazier}, {Brook}, {Burke-Spolaor}, {B{\'e}csy}, {Casey-Clyde},
  {Charisi}, {Chatterjee}, {Chen}, {Cordes}, {Cornish}, {Crawford},
  {Cromartie}, {DeCesar}, {Demorest}, {Dolch}, {Drachler}, {Ellis}, {Ferrara},
  {Fiore}, {Fonseca}, {Freedman}, {Garver-Daniels}, {Gentile}, {Glaser},
  {Good}, {G{\"u}ltekin}, {Hazboun}, {Jennings}, {Johnson}, {Jones}, {Kaiser},
  {Kaplan}, {Kelley}, {Shapiro Key}, {Laal}, {Lam}, {Lamb}, {Lazio},
  {Lewandowska}, {Liu}, {Lorimer}, {Luo}, {Lynch}, {Madison}, {McEwen},
  {McLaughlin}, {Mingarelli}, {Ng}, {Nice}, {Ocker}, {Olum}, {Pennucci}, {Pol},
  {Ransom}, {Ray}, {Romano}, {Shapiro-Albert}, {Siemens}, {Simon}, {Siwek},
  {Spiewak}, {Stairs}, {Stinebring}, {Stovall}, {Swiggum}, {Sydnor}, {Taylor},
  {Turner}, {Vallisneri}, {Vigeland}, {Wahl}, {Walsh}, {Witt}, \&
  {Young}}]{nanograv_12p5yr_cw}
{Arzoumanian}, Z., {Baker}, P.~T., {Blecha}, L., {et~al.} 2023, arXiv e-prints,
  arXiv:2301.03608, \dodoi{10.48550/arXiv.2301.03608}

\bibitem[{{Astropy Collaboration} {et~al.}(2013){Astropy Collaboration},
  {Robitaille}, {Tollerud}, {Greenfield}, {Droettboom}, {Bray}, {Aldcroft},
  {Davis}, {Ginsburg}, {Price-Whelan}, {Kerzendorf}, {Conley}, {Crighton},
  {Barbary}, {Muna}, {Ferguson}, {Grollier}, {Parikh}, {Nair}, {Unther},
  {Deil}, {Woillez}, {Conseil}, {Kramer}, {Turner}, {Singer}, {Fox}, {Weaver},
  {Zabalza}, {Edwards}, {Azalee Bostroem}, {Burke}, {Casey}, {Crawford},
  {Dencheva}, {Ely}, {Jenness}, {Labrie}, {Lim}, {Pierfederici}, {Pontzen},
  {Ptak}, {Refsdal}, {Servillat}, \& {Streicher}}]{astropy:2013}
{Astropy Collaboration}, {Robitaille}, T.~P., {Tollerud}, E.~J., {et~al.} 2013,
  \aap, 558, A33, \dodoi{10.1051/0004-6361/201322068}

\bibitem[{{Babak} \& {Sesana}(2012)}]{babak_sesana_multiple_cw}
{Babak}, S., \& {Sesana}, A. 2012, \prd, 85, 044034,
  \dodoi{10.1103/PhysRevD.85.044034}

\bibitem[{{Babak} {et~al.}(2016){Babak}, {Petiteau}, {Sesana}, {Brem},
  {Rosado}, {Taylor}, {Lassus}, {Hessels}, {Bassa}, {Burgay}, {Caballero},
  {Champion}, {Cognard}, {Desvignes}, {Gair}, {Guillemot}, {Janssen},
  {Karuppusamy}, {Kramer}, {Lazarus}, {Lee}, {Lentati}, {Liu}, {Mingarelli},
  {Os{\l}owski}, {Perrodin}, {Possenti}, {Purver}, {Sanidas}, {Smits},
  {Stappers}, {Theureau}, {Tiburzi}, {van Haasteren}, {Vecchio}, \&
  {Verbiest}}]{EPTA-CW-paper}
{Babak}, S., {Petiteau}, A., {Sesana}, A., {et~al.} 2016, \mnras, 455, 1665,
  \dodoi{10.1093/mnras/stv2092}

\bibitem[{{B{\'e}csy} \& {Cornish}(2020)}]{BayesHopper}
{B{\'e}csy}, B., \& {Cornish}, N.~J. 2020, Classical and Quantum Gravity, 37,
  135011, \dodoi{10.1088/1361-6382/ab8bbd}

\bibitem[{{B{\'e}csy} {et~al.}(2022){B{\'e}csy}, {Cornish}, \&
  {Digman}}]{QuickCW}
{B{\'e}csy}, B., {Cornish}, N.~J., \& {Digman}, M.~C. 2022, \prd, 105, 122003,
  \dodoi{10.1103/PhysRevD.105.122003}

\bibitem[{B\'ecsy {et~al.}(2022)B\'ecsy, Cornish, \& Kelley}]{Becsy:2022pnr}
B\'ecsy, B., Cornish, N.~J., \& Kelley, L.~Z. 2022, Astrophys. J., 941, 119,
  \dodoi{10.3847/1538-4357/aca1b2}

\bibitem[{B{\'e}csy {et~al.}(2023)B{\'e}csy, Digman, \& Cornish}]{QuickCW_code}
B{\'e}csy, B., Digman, M., \& Cornish, N.~J. 2023, QuickCW (v1.0.1).
\newblock \url{https://github.com/nanograv/QuickCW/tree/v1.0.1}

\bibitem[{{Begelman} {et~al.}(1980){Begelman}, {Blandford}, \&
  {Rees}}]{Begelman+1980}
{Begelman}, M.~C., {Blandford}, R.~D., \& {Rees}, M.~J. 1980, \nat, 287, 307,
  \dodoi{10.1038/287307a0}

\bibitem[{{Bogdanovi{\'c}} {et~al.}(2022){Bogdanovi{\'c}}, {Miller}, \&
  {Blecha}}]{2022LRR....25....3B}
{Bogdanovi{\'c}}, T., {Miller}, M.~C., \& {Blecha}, L. 2022, Living Reviews in
  Relativity, 25, 3, \dodoi{10.1007/s41114-022-00037-8}

\bibitem[{{Burke-Spolaor} {et~al.}(2019){Burke-Spolaor}, {Taylor}, {Charisi},
  {Dolch}, {Hazboun}, {Holgado}, {Kelley}, {Lazio}, {Madison}, {McMann},
  {Mingarelli}, {Rasskazov}, {Siemens}, {Simon}, \& {Smith}}]{PTA_review}
{Burke-Spolaor}, S., {Taylor}, S.~R., {Charisi}, M., {et~al.} 2019, \aapr, 27,
  5, \dodoi{10.1007/s00159-019-0115-7}

\bibitem[{{Chamberlin} {et~al.}(2015){Chamberlin}, {Creighton}, {Siemens},
  {Demorest}, {Ellis}, {Price}, \& {Romano}}]{os}
{Chamberlin}, S.~J., {Creighton}, J. D.~E., {Siemens}, X., {et~al.} 2015, \prd,
  91, 044048, \dodoi{10.1103/PhysRevD.91.044048}

\bibitem[{{Charisi} {et~al.}(2016){Charisi}, {Bartos}, {Haiman},
  {Price-Whelan}, {Graham}, {Bellm}, {Laher}, \& {Marka}}]{Charisi+2016}
{Charisi}, M., {Bartos}, I., {Haiman}, Z., {et~al.} 2016, ArXiv e-prints.
\newblock \doarXiv{1604.01020}

\bibitem[{{Charisi} {et~al.}(2022){Charisi}, {Taylor}, {Runnoe}, {Bogdanovic},
  \& {Trump}}]{Charisi+2022}
{Charisi}, M., {Taylor}, S.~R., {Runnoe}, J., {Bogdanovic}, T., \& {Trump},
  J.~R. 2022, \mnras, 510, 5929, \dodoi{10.1093/mnras/stab3713}

\bibitem[{{Chatterjee} {et~al.}(2009){Chatterjee}, {Brisken}, {Vlemmings},
  {Goss}, {Lazio}, {Cordes}, {Thorsett}, {Fomalont}, {Lyne}, \&
  {Kramer}}]{http://adsabs.harvard.edu/abs/2009ApJ...698..250C}
{Chatterjee}, S., {Brisken}, W.~F., {Vlemmings}, W.~H.~T., {et~al.} 2009, \apj,
  698, 250, \dodoi{10.1088/0004-637X/698/1/250}

\bibitem[{{Chen} {et~al.}(2020){Chen}, {Liu}, {Liao}, {Holgado}, {Guo},
  {Gruendl}, {Morganson}, {Shen}, {Zhang}, {Abbott}, {Aguena}, {Allam},
  {Avila}, {Bertin}, {Bhargava}, {Brooks}, {Burke}, {Carnero Rosell},
  {Carollo}, {Carrasco Kind}, {Carretero}, {Costanzi}, {da Costa}, {Davis}, {De
  Vicente}, {Desai}, {Diehl}, {Doel}, {Everett}, {Flaugher}, {Friedel},
  {Frieman}, {Garc{\'\i}a-Bellido}, {Gaztanaga}, {Glazebrook}, {Gruen},
  {Gutierrez}, {Hinton}, {Hollowood}, {James}, {Kim}, {Kuehn}, {Kuropatkin},
  {Lewis}, {Lidman}, {Lima}, {Maia}, {March}, {Marshall}, {Menanteau},
  {Miquel}, {Palmese}, {Paz-Chinch{\'o}n}, {Plazas}, {Sanchez}, {Schubnell},
  {Serrano}, {Sevilla-Noarbe}, {Smith}, {Suchyta}, {Swanson}, {Tarle},
  {Tucker}, {Norbert Varga}, \& {Walker}}]{Chen_DES_PLCs+2020}
{Chen}, Y.-C., {Liu}, X., {Liao}, W.-T., {et~al.} 2020, \mnras, 499, 2245,
  \dodoi{10.1093/mnras/staa2957}

\bibitem[{{Chen} {et~al.}(2022){Chen}, {Zhai}, {Liu}, {Guo}, {Peng}, {Li},
  {SongSheng}, {Du}, {Hu}, \& {Wang}}]{Chen_ZTF_PLCs+2022}
{Chen}, Y.-J., {Zhai}, S., {Liu}, J.-R., {et~al.} 2022, arXiv e-prints,
  arXiv:2206.11497.
\newblock \doarXiv{2206.11497}

\bibitem[{{Cordes} \& {Lazio}(2002)}]{ne2001}
{Cordes}, J.~M., \& {Lazio}, T.~J.~W. 2002, arXiv e-prints, astro,
  \dodoi{10.48550/arXiv.astro-ph/0207156}

\bibitem[{Cornish \& Sesana(2013)}]{Cornish:2013aba}
Cornish, N.~J., \& Sesana, A. 2013, Class. Quant. Grav., 30, 224005,
  \dodoi{10.1088/0264-9381/30/22/224005}

\bibitem[{{De Rosa} {et~al.}(2019){De Rosa}, {Vignali}, {Bogdanovi{\'c}},
  {Capelo}, {Charisi}, {Dotti}, {Husemann}, {Lusso}, {Mayer}, {Paragi},
  {Runnoe}, {Sesana}, {Steinborn}, {Bianchi}, {Colpi}, {del Valle}, {Frey},
  {Gab{\'a}nyi}, {Giustini}, {Guainazzi}, {Haiman}, {Herrera Ruiz},
  {Herrero-Illana}, {Iwasawa}, {Komossa}, {Lena}, {Loiseau}, {Perez-Torres},
  {Piconcelli}, \& {Volonteri}}]{DeRosa+2019}
{De Rosa}, A., {Vignali}, C., {Bogdanovi{\'c}}, T., {et~al.} 2019, \nar, 86,
  101525, \dodoi{10.1016/j.newar.2020.101525}

\bibitem[{{Deller} {et~al.}(2008){Deller}, {Verbiest}, {Tingay}, \&
  {Bailes}}]{http://adsabs.harvard.edu/abs/2008ApJ...685L..67D}
{Deller}, A.~T., {Verbiest}, J.~P.~W., {Tingay}, S.~J., \& {Bailes}, M. 2008,
  \apjl, 685, L67, \dodoi{10.1086/592401}

\bibitem[{{Deller} {et~al.}(2019{\natexlab{a}}){Deller}, {Goss}, {Brisken},
  {Chatterjee}, {Cordes}, {Janssen}, {Kovalev}, {Lazio}, {Petrov}, {Stappers},
  \& {Lyne}}]{https://ui.adsabs.harvard.edu/abs/2019ApJ...875..100D}
{Deller}, A.~T., {Goss}, W.~M., {Brisken}, W.~F., {et~al.} 2019{\natexlab{a}},
  \apj, 875, 100, \dodoi{10.3847/1538-4357/ab11c7}

\bibitem[{{Deller} {et~al.}(2019{\natexlab{b}}){Deller}, {Goss}, {Brisken},
  {Chatterjee}, {Cordes}, {Janssen}, {Kovalev}, {Lazio}, {Petrov}, {Stappers},
  \& {Lyne}}]{https://ui.adsabs.harvard.edu/abs/2019ApJ...875..100D/}
---. 2019{\natexlab{b}}, \apj, 875, 100, \dodoi{10.3847/1538-4357/ab11c7}

\bibitem[{Dickey(1971)}]{SD_BF}
Dickey, J.~M. 1971, The Annals of Mathematical Statistics, 42, 204.
\newblock \url{http://www.jstor.org/stable/2958475}

\bibitem[{{Ding} {et~al.}(2020){Ding}, {Deller}, {Freire}, {Kaplan}, {Lazio},
  {Shannon}, \&
  {Stappers}}]{ttps://ui.adsabs.harvard.edu/abs/2020ApJ...896...85D}
{Ding}, H., {Deller}, A.~T., {Freire}, P., {et~al.} 2020, \apj, 896, 85,
  \dodoi{10.3847/1538-4357/ab8f27}

\bibitem[{{Ding} {et~al.}(2023){Ding}, {Deller}, {Stappers}, {Lazio}, {Kaplan},
  {Chatterjee}, {Brisken}, {Cordes}, {Freire}, {Fonseca}, {Stairs},
  {Guillemot}, {Lyne}, {Cognard}, {Reardon}, \&
  {Theureau}}]{https://doi.org/10.1093/mnras/stac3725}
{Ding}, H., {Deller}, A.~T., {Stappers}, B.~W., {et~al.} 2023, \mnras, 519,
  4982, \dodoi{10.1093/mnras/stac3725}

\bibitem[{{Ellis} {et~al.}(2019){Ellis}, {Vallisneri}, {Taylor}, \&
  {Baker}}]{enterprise}
{Ellis}, J.~A., {Vallisneri}, M., {Taylor}, S.~R., \& {Baker}, P.~T. 2019,
  {ENTERPRISE: Enhanced Numerical Toolbox Enabling a Robust PulsaR Inference
  SuitE}.
\newblock \doeprint{1912.015}

\bibitem[{{Falxa} {et~al.}(2023){Falxa}, {Babak}, {Baker}, {B{\'e}csy},
  {Chalumeau}, {Chen}, {Chen}, {Cornish}, {Guillemot}, {Hazboun}, {Mingarelli},
  {Parthasarathy}, {Petiteau}, {Pol}, {Sesana}, {Spolaor}, {Taylor},
  {Theureau}, {Vallisneri}, {Vigeland}, {Witt}, {Zhu}, {Antoniadis},
  {Arzoumanian}, {Bailes}, {Bhat}, {Blecha}, {Brazier}, {Brook}, {Caballero},
  {Cameron}, {Casey-Clyde}, {Champion}, {Charisi}, {Chatterjee}, {Cognard},
  {Cordes}, {Crawford}, {Cromartie}, {Crowter}, {Dai}, {DeCesar}, {Demorest},
  {Desvignes}, {Dolch}, {Drachler}, {Feng}, {Ferrara}, {Fiore}, {Fonseca},
  {Garver-Daniels}, {Glaser}, {Goncharov}, {Good}, {Griessmeier}, {Guo},
  {G{\"u}ltekin}, {Hobbs}, {Hu}, {Islo}, {Jang}, {Jennings}, {Johnson},
  {Jones}, {Kaczmarek}, {Kaiser}, {Kaplan}, {Keith}, {Kelley}, {Kerr}, {Key},
  {Laal}, {Lam}, {Lamb}, {Lazio}, {Liu}, {Liu}, {Luo}, {Lynch}, {Madison},
  {Main}, {Manchester}, {McEwen}, {McKee}, {McLaughlin}, {Ng}, {Nice}, {Ocker},
  {Olum}, {Os{\l}owski}, {Pennucci}, {Perera}, {Perrodin}, {Porayko},
  {Possenti}, {Quelquejay-Leclere}, {Ransom}, {Ray}, {Reardon}, {Russell},
  {Samajdar}, {Sarkissian}, {Schult}, {Shaifullah}, {Shannon},
  {Shapiro-Albert}, {Siemens}, {Simon}, {Siwek}, {Smith}, {Speri}, {Spiewak},
  {Stairs}, {Stappers}, {Stinebring}, {Swiggum}, {Tiburzi}, {Turner},
  {Vecchio}, {Verbiest}, {Wahl}, {Wang}, {Wang}, {Wang}, {Wu}, {Zhang}, \&
  {Zhang}}]{ipta_dr2_cw}
{Falxa}, M., {Babak}, S., {Baker}, P.~T., {et~al.} 2023, \mnras,
  \dodoi{10.1093/mnras/stad812}

\bibitem[{{Freire} {et~al.}(2012){Freire}, {Wex}, {Esposito-Far{\`e}se},
  {Verbiest}, {Bailes}, {Jacoby}, {Kramer}, {Stairs}, {Antoniadis}, \&
  {Janssen}}]{http://adsabs.harvard.edu/abs/2012MNRAS.423.3328F}
{Freire}, P. C.~C., {Wex}, N., {Esposito-Far{\`e}se}, G., {et~al.} 2012,
  \mnras, 423, 3328, \dodoi{10.1111/j.1365-2966.2012.21253.x}

\bibitem[{{Goncharov} {et~al.}(2021){Goncharov}, {Reardon}, {Shannon}, {Zhu},
  {Thrane}, {Bailes}, {Bhat}, {Dai}, {Hobbs}, {Kerr}, {Manchester},
  {Os{\l}owski}, {Parthasarathy}, {Russell}, {Spiewak}, {Thyagarajan}, \&
  {Wang}}]{ppta_dr2_noise}
{Goncharov}, B., {Reardon}, D.~J., {Shannon}, R.~M., {et~al.} 2021, \mnras,
  502, 478, \dodoi{10.1093/mnras/staa3411}

\bibitem[{{G{\'o}rski} {et~al.}(2005){G{\'o}rski}, {Hivon}, {Banday},
  {Wandelt}, {Hansen}, {Reinecke}, \& {Bartelmann}}]{healpix}
{G{\'o}rski}, K.~M., {Hivon}, E., {Banday}, A.~J., {et~al.} 2005, \apj, 622,
  759, \dodoi{10.1086/427976}

\bibitem[{{Graham} {et~al.}(2015{\natexlab{a}}){Graham}, {Djorgovski}, {Stern},
  {Glikman}, {Drake}, {Mahabal}, {Donalek}, {Larson}, \&
  {Christensen}}]{Graham+2015}
{Graham}, M.~J., {Djorgovski}, S.~G., {Stern}, D., {et~al.} 2015{\natexlab{a}},
  \nat, 518, 74, \dodoi{10.1038/nature14143}

\bibitem[{{Graham} {et~al.}(2015{\natexlab{b}}){Graham}, {Djorgovski}, {Stern},
  {Drake}, {Mahabal}, {Donalek}, {Glikman}, {Larson}, \&
  {Christensen}}]{Graham+2015b}
---. 2015{\natexlab{b}}, \mnras, 453, 1562, \dodoi{10.1093/mnras/stv1726}

\bibitem[{{Guillemot} {et~al.}(2016){Guillemot}, {Smith}, {Laffon}, {Janssen},
  {Cognard}, {Theureau}, {Desvignes}, {Ferrara}, \&
  {Ray}}]{http://dx.doi.org/10.1051/0004-6361/201527847}
{Guillemot}, L., {Smith}, D.~A., {Laffon}, H., {et~al.} 2016, \aap, 587, A109,
  \dodoi{10.1051/0004-6361/201527847}

\bibitem[{{Hazboun} {et~al.}(2019){Hazboun}, {Romano}, \&
  {Smith}}]{jeff_sensitivity}
{Hazboun}, J.~S., {Romano}, J.~D., \& {Smith}, T.~L. 2019, \prd, 100, 104028,
  \dodoi{10.1103/PhysRevD.100.104028}

\bibitem[{{Hazboun} {et~al.}(2020){Hazboun}, {Simon}, {Taylor}, {Lam},
  {Vigeland}, {Islo}, {Key}, {Arzoumanian}, {Baker}, {Brazier}, {Brook},
  {Burke-Spolaor}, {Chatterjee}, {Cordes}, {Cornish}, {Crawford}, {Crowter},
  {Cromartie}, {DeCesar}, {Demorest}, {Dolch}, {Ellis}, {Ferdman}, {Ferrara},
  {Fonseca}, {Garver-Daniels}, {Gentile}, {Good}, {Holgado}, {Huerta},
  {Jennings}, {Jones}, {Jones}, {Kaiser}, {Kaplan}, {Kelley}, {Lazio}, {Levin},
  {Lommen}, {Lorimer}, {Luo}, {Lynch}, {Madison}, {McLaughlin}, {McWilliams},
  {Mingarelli}, {Ng}, {Nice}, {Pennucci}, {Pol}, {Ransom}, {Ray}, {Siemens},
  {Spiewak}, {Stairs}, {Stinebring}, {Stovall}, {Swiggum}, {Turner},
  {Vallisneri}, {van Haasteren}, {Witt}, \& {Zhu}}]{11yr_slice}
{Hazboun}, J.~S., {Simon}, J., {Taylor}, S.~R., {et~al.} 2020, \apj, 890, 108,
  \dodoi{10.3847/1538-4357/ab68db}

\bibitem[{{Hellings} \& {Downs}(1983)}]{HD}
{Hellings}, R.~W., \& {Downs}, G.~S. 1983, \apjl, 265, L39,
  \dodoi{10.1086/183954}

\bibitem[{{Hinshaw} {et~al.}(2013){Hinshaw}, {Larson}, {Komatsu}, {Spergel},
  {Bennett}, {Dunkley}, {Nolta}, {Halpern}, {Hill}, {Odegard}, {Page}, {Smith},
  {Weiland}, {Gold}, {Jarosik}, {Kogut}, {Limon}, {Meyer}, {Tucker}, {Wollack},
  \& {Wright}}]{wmap9}
{Hinshaw}, G., {Larson}, D., {Komatsu}, E., {et~al.} 2013, \apjs, 208, 19,
  \dodoi{10.1088/0067-0049/208/2/19}

\bibitem[{{Hobbs} {et~al.}(2006){Hobbs}, {Edwards}, \& {Manchester}}]{tempo2}
{Hobbs}, G.~B., {Edwards}, R.~T., \& {Manchester}, R.~N. 2006, \mnras, 369,
  655, \dodoi{10.1111/j.1365-2966.2006.10302.x}

\bibitem[{{Hotan} {et~al.}(2004){Hotan}, {Bailes}, \&
  {Ord}}]{http://adsabs.harvard.edu/abs/2004MNRAS.355..941H}
{Hotan}, A.~W., {Bailes}, M., \& {Ord}, S.~M. 2004, \mnras, 355, 941,
  \dodoi{10.1111/j.1365-2966.2004.08376.x}

\bibitem[{{Hotan} {et~al.}(2006){Hotan}, {Bailes}, \&
  {Ord}}]{http://adsabs.harvard.edu/abs/2006MNRAS.369.1502H}
---. 2006, \mnras, 369, 1502, \dodoi{10.1111/j.1365-2966.2006.10394.x}

\bibitem[{{Hourihane} {et~al.}(2022){Hourihane}, {Meyers}, {Johnson},
  {Chatziioannou}, \& {Vallisneri}}]{sophie_resampling}
{Hourihane}, S., {Meyers}, P., {Johnson}, A., {Chatziioannou}, K., \&
  {Vallisneri}, M. 2022, arXiv e-prints, arXiv:2212.06276,
  \dodoi{10.48550/arXiv.2212.06276}

\bibitem[{Hunter(2007)}]{matplotlib}
Hunter, J.~D. 2007, Computing in Science \& Engineering, 9, 90,
  \dodoi{10.1109/MCSE.2007.55}

\bibitem[{{Jacoby} {et~al.}(2005){Jacoby}, {Hotan}, {Bailes}, {Ord}, \&
  {Kulkarni}}]{http://adsabs.harvard.edu/abs/2005ApJ...629L.113J}
{Jacoby}, B.~A., {Hotan}, A., {Bailes}, M., {Ord}, S., \& {Kulkarni}, S.~R.
  2005, \apjl, 629, L113, \dodoi{10.1086/449311}

\bibitem[{{Jenet} {et~al.}(2004){Jenet}, {Lommen}, {Larson}, \&
  {Wen}}]{Jenet+2004}
{Jenet}, F.~A., {Lommen}, A., {Larson}, S.~L., \& {Wen}, L. 2004, \apj, 606,
  799, \dodoi{10.1086/383020}

\bibitem[{{Jennings} {et~al.}(2018){Jennings}, {Kaplan}, {Chatterjee},
  {Cordes}, \&
  {Deller}}]{ttps://ui.adsabs.harvard.edu/abs/2018ApJ...864...26J/}
{Jennings}, R.~J., {Kaplan}, D.~L., {Chatterjee}, S., {Cordes}, J.~M., \&
  {Deller}, A.~T. 2018, \apj, 864, 26, \dodoi{10.3847/1538-4357/aad084}

\bibitem[{{Johnson} {et~al.}(2023)}]{code_review}
{Johnson}, A., {et~al.} 2023, in preparation

\bibitem[{{Kaiser} {et~al.}(2023)}]{bayesian_timing}
{Kaiser}, A.~R., {et~al.} 2023, in preparation

\bibitem[{{Kaspi} {et~al.}(1994){Kaspi}, {Taylor}, \&
  {Ryba}}]{http://adsabs.harvard.edu/abs/1994ApJ...428..713K}
{Kaspi}, V.~M., {Taylor}, J.~H., \& {Ryba}, M.~F. 1994, \apj, 428, 713,
  \dodoi{10.1086/174280}

\bibitem[{{Kelley} {et~al.}(2019){Kelley}, {Charisi}, {Burke-Spolaor}, {Simon},
  {Blecha}, {Bogdanovic}, {Colpi}, {Comerford}, {D'Orazio}, {Dotti},
  {Eracleous}, {Graham}, {Greene}, {Haiman}, {Holley-Bockelmann}, {Kara},
  {Kelly}, {Komossa}, {Larson}, {Liu}, {Ma}, {Noble}, {Paschalidis}, {Rafikov},
  {Ravi}, {Runnoe}, {Sesana}, {Stern}, {Strauss}, {U}, {Volonteri}, \&
  {Nanograv Collaboration}}]{Kelley_white_paper_2019}
{Kelley}, L., {Charisi}, M., {Burke-Spolaor}, S., {et~al.} 2019, \baas, 51,
  490, \dodoi{10.48550/arXiv.1903.07644}

\bibitem[{Kelley(2022)}]{cosmopy}
Kelley, L.~Z. 2022, cosmopy (v3.5.5).
\newblock \url{https://github.com/lzkelley/cosmopy/tree/v3.5.5}

\bibitem[{{Kelley} {et~al.}(2018){Kelley}, {Blecha}, {Hernquist}, {Sesana}, \&
  {Taylor}}]{Luke_single_source}
{Kelley}, L.~Z., {Blecha}, L., {Hernquist}, L., {Sesana}, A., \& {Taylor},
  S.~R. 2018, \mnras, 477, 964, \dodoi{10.1093/mnras/sty689}

\bibitem[{{Komossa} {et~al.}(2023){Komossa}, {Kraus}, {Grupe}, {Gonzalez},
  {Gurwell}, {Gallo}, {Liu}, {Myserlis}, {Krichbaum}, {Laine}, {Bach},
  {G{\'o}mez}, {Parker}, {Yao}, \& {Berton}}]{Komossa+2023}
{Komossa}, S., {Kraus}, A., {Grupe}, D., {et~al.} 2023, \apj, 944, 177,
  \dodoi{10.3847/1538-4357/acaf71}

\bibitem[{{Kormendy} \& {Ho}(2013)}]{Kormendy+2013}
{Kormendy}, J., \& {Ho}, L.~C. 2013, \araa, 51, 511,
  \dodoi{10.1146/annurev-astro-082708-101811}

\bibitem[{{Lazaridis} {et~al.}(2009){Lazaridis}, {Wex}, {Jessner}, {Kramer},
  {Stappers}, {Janssen}, {Desvignes}, {Purver}, {Cognard}, {Theureau}, {Lyne},
  {Jordan}, \& {Zensus}}]{http://adsabs.harvard.edu/abs/2009MNRAS.400..805L}
{Lazaridis}, K., {Wex}, N., {Jessner}, A., {et~al.} 2009, \mnras, 400, 805,
  \dodoi{10.1111/j.1365-2966.2009.15481.x}

\bibitem[{{Liu} {et~al.}(2023){Liu}, {Cohen}, {McGrath}, {Demorest}, \&
  {Vigeland}}]{tingting2023}
{Liu}, T., {Cohen}, T., {McGrath}, C., {Demorest}, P.~B., \& {Vigeland}, S.~J.
  2023, \apj, 945, 78, \dodoi{10.3847/1538-4357/acb492}

\bibitem[{{Liu} \& {Vigeland}(2021)}]{tingting2021}
{Liu}, T., \& {Vigeland}, S.~J. 2021, \apj, 921, 178,
  \dodoi{10.3847/1538-4357/ac1da9}

\bibitem[{{Liu} {et~al.}(2019){Liu}, {Gezari}, {Ayers}, {Burgett}, {Chambers},
  {Hodapp}, {Huber}, {Kudritzki}, {Metcalfe}, {Tonry}, {Wainscoat}, \&
  {Waters}}]{LiuGezari+2019}
{Liu}, T., {Gezari}, S., {Ayers}, M., {et~al.} 2019, \apj, 884, 36,
  \dodoi{10.3847/1538-4357/ab40cb}

\bibitem[{{Lommen} {et~al.}(2006){Lommen}, {Kipphorn}, {Nice}, {Splaver},
  {Stairs}, \& {Backer}}]{http://adsabs.harvard.edu/abs/2006ApJ...642.1012L}
{Lommen}, A.~N., {Kipphorn}, R.~A., {Nice}, D.~J., {et~al.} 2006, \apj, 642,
  1012, \dodoi{10.1086/501067}

\bibitem[{{Luo} {et~al.}(2019){Luo}, {Ransom}, {Demorest}, {van Haasteren},
  {Ray}, {Stovall}, {Bachetti}, {Archibald}, {Kerr}, {Colen}, \&
  {Jenet}}]{pint}
{Luo}, J., {Ransom}, S., {Demorest}, P., {et~al.} 2019, {PINT: High-precision
  pulsar timing analysis package}, Astrophysics Source Code Library, record
  ascl:1902.007.
\newblock \doeprint{1902.007}

\bibitem[{{Meyers} {et~al.}(2023){Meyers}, {Chatziioannou}, {Vallisneri}, \&
  {Chua}}]{meyers+23}
{Meyers}, P.~M., {Chatziioannou}, K., {Vallisneri}, M., \& {Chua}, A. J.~K.
  2023, arXiv e-prints, arXiv:2306.05559, \dodoi{10.48550/arXiv.2306.05559}

\bibitem[{Mingarelli {et~al.}(2013)Mingarelli, Sidery, Mandel, \&
  Vecchio}]{Mingarelli:2013dsa}
Mingarelli, C. M.~F., Sidery, T., Mandel, I., \& Vecchio, A. 2013, Phys. Rev.
  D, 88, 062005, \dodoi{10.1103/PhysRevD.88.062005}

\bibitem[{{Mingarelli} {et~al.}(2017){Mingarelli}, {Lazio}, {Sesana}, {Greene},
  {Ellis}, {Ma}, {Croft}, {Burke-Spolaor}, \& {Taylor}}]{Mingarelli:2017}
{Mingarelli}, C. M.~F., {Lazio}, T. J.~W., {Sesana}, A., {et~al.} 2017, Nature
  Astronomy, 1, 886, \dodoi{10.1038/s41550-017-0299-6}

\bibitem[{{NANOGrav Collaboration}(2023)}]{nanograv_15yr_cw_sup_mat}
{NANOGrav Collaboration}. 2023, {CW analysis of the NANOGrav 15-year dataset},
  1.0,  Zenodo, \dodoi{10.5281/zenodo.8067506}

\bibitem[{{Nice} {et~al.}(2015){Nice}, {Demorest}, {Stairs}, {Manchester},
  {Taylor}, {Peters}, {Weisberg}, {Irwin}, {Wex}, \& {Huang}}]{tempo}
{Nice}, D., {Demorest}, P., {Stairs}, I., {et~al.} 2015, {Tempo: Pulsar timing
  data analysis}, Astrophysics Source Code Library, record ascl:1509.002.
\newblock \doeprint{1509.002}

\bibitem[{{Price-Whelan} {et~al.}(2018){Price-Whelan}, {Sip{\H{o}}cz},
  {G{\"u}nther}, {Lim}, {Crawford}, {Conseil}, {Shupe}, {Craig}, {Dencheva},
  {Ginsburg}, {VanderPlas}, {Bradley}, {P{\'e}rez-Su{\'a}rez}, {de Val-Borro},
  {Paper Contributors}, {Aldcroft}, {Cruz}, {Robitaille}, {Tollerud},
  {Coordination Committee}, {Ardelean}, {Babej}, {Bach}, {Bachetti}, {Bakanov},
  {Bamford}, {Barentsen}, {Barmby}, {Baumbach}, {Berry}, {Biscani}, {Boquien},
  {Bostroem}, {Bouma}, {Brammer}, {Bray}, {Breytenbach}, {Buddelmeijer},
  {Burke}, {Calderone}, {Cano Rodr{\'\i}guez}, {Cara}, {Cardoso}, {Cheedella},
  {Copin}, {Corrales}, {Crichton}, {D{\textquoteright}Avella}, {Deil},
  {Depagne}, {Dietrich}, {Donath}, {Droettboom}, {Earl}, {Erben}, {Fabbro},
  {Ferreira}, {Finethy}, {Fox}, {Garrison}, {Gibbons}, {Goldstein}, {Gommers},
  {Greco}, {Greenfield}, {Groener}, {Grollier}, {Hagen}, {Hirst}, {Homeier},
  {Horton}, {Hosseinzadeh}, {Hu}, {Hunkeler}, {Ivezi{\'c}}, {Jain}, {Jenness},
  {Kanarek}, {Kendrew}, {Kern}, {Kerzendorf}, {Khvalko}, {King}, {Kirkby},
  {Kulkarni}, {Kumar}, {Lee}, {Lenz}, {Littlefair}, {Ma}, {Macleod},
  {Mastropietro}, {McCully}, {Montagnac}, {Morris}, {Mueller}, {Mumford},
  {Muna}, {Murphy}, {Nelson}, {Nguyen}, {Ninan}, {N{\"o}the}, {Ogaz}, {Oh},
  {Parejko}, {Parley}, {Pascual}, {Patil}, {Patil}, {Plunkett}, {Prochaska},
  {Rastogi}, {Reddy Janga}, {Sabater}, {Sakurikar}, {Seifert}, {Sherbert},
  {Sherwood-Taylor}, {Shih}, {Sick}, {Silbiger}, {Singanamalla}, {Singer},
  {Sladen}, {Sooley}, {Sornarajah}, {Streicher}, {Teuben}, {Thomas},
  {Tremblay}, {Turner}, {Terr{\'o}n}, {van Kerkwijk}, {de la Vega}, {Watkins},
  {Weaver}, {Whitmore}, {Woillez}, {Zabalza}, \& {Contributors}}]{astropy}
{Price-Whelan}, A.~M., {Sip{\H{o}}cz}, B.~M., {G{\"u}nther}, H.~M., {et~al.}
  2018, \aj, 156, 123, \dodoi{10.3847/1538-3881/aabc4f}

\bibitem[{{Reardon} {et~al.}(2023)}]{ppta_dr3_gwb}
{Reardon}, D.~J., {et~al.} 2023, TBD

\bibitem[{{Rosado} {et~al.}(2015){Rosado}, {Sesana}, \&
  {Gair}}]{rosado_expected_properties}
{Rosado}, P.~A., {Sesana}, A., \& {Gair}, J. 2015, \mnras, 451, 2417,
  \dodoi{10.1093/mnras/stv1098}

\bibitem[{{Sandhu} {et~al.}(1997){Sandhu}, {Bailes}, {Manchester}, {Navarro},
  {Kulkarni}, \&
  {Anderson}}]{http://adsabs.harvard.edu/abs/1997ApJ...478L..95S}
{Sandhu}, J.~S., {Bailes}, M., {Manchester}, R.~N., {et~al.} 1997, \apjl, 478,
  L95, \dodoi{10.1086/310562}

\bibitem[{{Sardesai} \& {Vigeland}(2023)}]{mcos}
{Sardesai}, S.~C., \& {Vigeland}, S.~J. 2023, arXiv e-prints, arXiv:2303.09615,
  \dodoi{10.48550/arXiv.2303.09615}

\bibitem[{{Schutz} \& {Ma}(2016)}]{Schutz+2016}
{Schutz}, K., \& {Ma}, C.-P. 2016, \mnras, 459, 1737,
  \dodoi{10.1093/mnras/stw768}

\bibitem[{{Sesana} {et~al.}(2009){Sesana}, {Vecchio}, \& {Volonteri}}]{SVV2009}
{Sesana}, A., {Vecchio}, A., \& {Volonteri}, M. 2009, \mnras, 394, 2255,
  \dodoi{10.1111/j.1365-2966.2009.14499.x}

\bibitem[{{Simon} {et~al.}(2023)}]{adv_noise}
{Simon}, J., {et~al.} 2023, in preparation

\bibitem[{{Splaver} {et~al.}(2005){Splaver}, {Nice}, {Stairs}, {Lommen}, \&
  {Backer}}]{http://adsabs.harvard.edu/abs/2005ApJ...620..405S}
{Splaver}, E.~M., {Nice}, D.~J., {Stairs}, I.~H., {Lommen}, A.~N., \& {Backer},
  D.~C. 2005, \apj, 620, 405, \dodoi{10.1086/426804}

\bibitem[{{Stovall} {et~al.}(2014){Stovall}, {Lynch}, {Ransom}, {Archibald},
  {Banaszak}, {Biwer}, {Boyles}, {Dartez}, {Day}, {Ford}, {Flanigan}, {Garcia},
  {Hessels}, {Hinojosa}, {Jenet}, {Kaplan}, {Karako-Argaman}, {Kaspi},
  {Kondratiev}, {Leake}, {Lorimer}, {Lunsford}, {Martinez}, {Mata},
  {McLaughlin}, {Roberts}, {Rohr}, {Siemens}, {Stairs}, {van Leeuwen},
  {Walker}, \& {Wells}}]{http://adsabs.harvard.edu/abs/2014ApJ...791...67S}
{Stovall}, K., {Lynch}, R.~S., {Ransom}, S.~M., {et~al.} 2014, \apj, 791, 67,
  \dodoi{10.1088/0004-637X/791/1/67}

\bibitem[{{Susobhanan}(2022)}]{abhimanyu_eccbin_2022}
{Susobhanan}, A. 2022, arXiv e-prints, arXiv:2210.11454,
  \dodoi{10.48550/arXiv.2210.11454}

\bibitem[{{Susobhanan} {et~al.}(2020){Susobhanan}, {Gopakumar}, {Hobbs}, \&
  {Taylor}}]{abhimanyu_eccbin_2020}
{Susobhanan}, A., {Gopakumar}, A., {Hobbs}, G., \& {Taylor}, S.~R. 2020, \prd,
  101, 043022, \dodoi{10.1103/PhysRevD.101.043022}

\bibitem[{{Tarafdar} {et~al.}(2022){Tarafdar}, {Nobleson}, {Rana}, {Singha},
  {Krishnakumar}, {Joshi}, {Paladi}, {Kolhe}, {Batra}, {Agarwal}, {Bathula},
  {Dandapat}, {Desai}, {Dey}, {Hisano}, {Ingale}, {Kato}, {Kharbanda},
  {Kikunaga}, {Marmat}, {Pandian}, {Prabu}, {Srivastava}, {Surnis}, {Susarla},
  {Susobhanan}, {Takahashi}, {Arumugam}, {Bagchi}, {Banik}, {De}, {Girgaonkar},
  {Gopakumar}, {Gupta}, {Maan}, {Manoharan}, {Naidu}, \&
  {Pathak}}]{inpta_dr1_data}
{Tarafdar}, P., {Nobleson}, K., {Rana}, P., {et~al.} 2022, \pasa, 39, e053,
  \dodoi{10.1017/pasa.2022.46}

\bibitem[{{Taylor}(2021)}]{SteveBook}
{Taylor}, S.~R. 2021, arXiv e-prints, arXiv:2105.13270.
\newblock \doarXiv{2105.13270}

\bibitem[{Taylor {et~al.}(2021)Taylor, Baker, Hazboun, Simon, \& Vigeland}]{ee}
Taylor, S.~R., Baker, P.~T., Hazboun, J.~S., Simon, J., \& Vigeland, S.~J.
  2021, enterprise\textunderscore extensions.
\newblock \url{https://github.com/nanograv/enterprise_extensions}

\bibitem[{{Taylor} {et~al.}(2016){Taylor}, {Huerta}, {Gair}, \&
  {McWilliams}}]{steve_eccbin}
{Taylor}, S.~R., {Huerta}, E.~A., {Gair}, J.~R., \& {McWilliams}, S.~T. 2016,
  \apj, 817, 70, \dodoi{10.3847/0004-637X/817/1/70}

\bibitem[{{Toscano} {et~al.}(1999){Toscano}, {Britton}, {Manchester}, {Bailes},
  {Sandhu}, {Kulkarni}, \&
  {Anderson}}]{http://adsabs.harvard.edu/abs/1999ApJ...523L.171T}
{Toscano}, M., {Britton}, M.~C., {Manchester}, R.~N., {et~al.} 1999, \apjl,
  523, L171, \dodoi{10.1086/312276}

\bibitem[{{Vallisneri}(2020)}]{libstempo}
{Vallisneri}, M. 2020, {libstempo: Python wrapper for Tempo2}, Astrophysics
  Source Code Library, record ascl:2002.017.
\newblock \doeprint{2002.017}

\bibitem[{{Valtonen} {et~al.}(2023){Valtonen}, {Zola}, {Gopakumar},
  {L{\"a}hteenm{\"a}ki}, {Tornikoski}, {Dey}, {Gupta}, {Pursimo}, {Knudstrup},
  {Gomez}, {Hudec}, {Jel{\'\i}nek}, {{\v{S}}trobl}, {Berdyugin}, {Ciprini},
  {Reichart}, {Kouprianov}, {Matsumoto}, {Drozdz}, {Mugrauer}, {Sadun},
  {Zejmo}, {Sillanp{\"a}{\"a}}, {Lehto}, {Nilsson}, {Imazawa}, \&
  {Uemura}}]{Valtonen+2023}
{Valtonen}, M.~J., {Zola}, S., {Gopakumar}, A., {et~al.} 2023, \mnras, 521,
  6143, \dodoi{10.1093/mnras/stad922}

\bibitem[{{van Straten} {et~al.}(2001){van Straten}, {Bailes}, {Britton},
  {Kulkarni}, {Anderson}, {Manchester}, \&
  {Sarkissian}}]{http://adsabs.harvard.edu/abs/2001Natur.412..158V}
{van Straten}, W., {Bailes}, M., {Britton}, M., {et~al.} 2001, \nat, 412, 158,
  \dodoi{10.1038/35084015}

\bibitem[{{Verbiest} {et~al.}(2008){Verbiest}, {Bailes}, {van Straten},
  {Hobbs}, {Edwards}, {Manchester}, {Bhat}, {Sarkissian}, {Jacoby}, \&
  {Kulkarni}}]{http://adsabs.harvard.edu/abs/2008ApJ...679..675V}
{Verbiest}, J.~P.~W., {Bailes}, M., {van Straten}, W., {et~al.} 2008, \apj,
  679, 675, \dodoi{10.1086/529576}

\bibitem[{{Verbiest} {et~al.}(2009){Verbiest}, {Bailes}, {Coles}, {Hobbs}, {van
  Straten}, {Champion}, {Jenet}, {Manchester}, {Bhat}, {Sarkissian}, {Yardley},
  {Burke-Spolaor}, {Hotan}, \&
  {You}}]{http://adsabs.harvard.edu/abs/2009MNRAS.400..951V}
{Verbiest}, J.~P.~W., {Bailes}, M., {Coles}, W.~A., {et~al.} 2009, \mnras, 400,
  951, \dodoi{10.1111/j.1365-2966.2009.15508.x}

\bibitem[{{Vigeland} {et~al.}(2018){Vigeland}, {Islo}, {Taylor}, \&
  {Ellis}}]{nmos}
{Vigeland}, S.~J., {Islo}, K., {Taylor}, S.~R., \& {Ellis}, J.~A. 2018, \prd,
  98, 044003, \dodoi{10.1103/PhysRevD.98.044003}

\bibitem[{{Xin} {et~al.}(2021){Xin}, {Mingarelli}, \&
  {Hazboun}}]{Xin_Mingarelli+}
{Xin}, C., {Mingarelli}, C. M.~F., \& {Hazboun}, J.~S. 2021, \apj, 915, 97,
  \dodoi{10.3847/1538-4357/ac01c5}

\bibitem[{{Yardley} {et~al.}(2010){Yardley}, {Hobbs}, {Jenet}, {Verbiest},
  {Wen}, {Manchester}, {Coles}, {van Straten}, {Bailes}, {Bhat},
  {Burke-Spolaor}, {Champion}, {Hotan}, \& {Sarkissian}}]{yardley+10}
{Yardley}, D.~R.~B., {Hobbs}, G.~B., {Jenet}, F.~A., {et~al.} 2010, \mnras,
  407, 669, \dodoi{10.1111/j.1365-2966.2010.16949.x}

\bibitem[{{Zhu} {et~al.}(2014){Zhu}, {Hobbs}, {Wen}, {Coles}, {Wang},
  {Shannon}, {Manchester}, {Bailes}, {Bhat}, {Burke-Spolaor}, {Dai}, {Keith},
  {Kerr}, {Levin}, {Madison}, {Os{\l}owski}, {Ravi}, {Toomey}, \& {van
  Straten}}]{PPTA-cw-paper}
{Zhu}, X.-J., {Hobbs}, G., {Wen}, L., {et~al.} 2014, \mnras, 444, 3709,
  \dodoi{10.1093/mnras/stu1717}

\bibitem[{{Zic} {et~al.}(2023)}]{ppta_dr3_dataset}
{Zic}, A., {et~al.} 2023, TBD

\bibitem[{Zonca {et~al.}(2019)Zonca, Singer, Lenz, Reinecke, Rosset, Hivon, \&
  Gorski}]{healpy}
Zonca, A., Singer, L., Lenz, D., {et~al.} 2019, Journal of Open Source
  Software, 4, 1298, \dodoi{10.21105/joss.01298}

\end{thebibliography}
\end{document}